\documentclass[twocolumn,showpacs,preprintnumbers,amsmath,amssymb, prb]{revtex4}

\usepackage{graphicx}
\usepackage{dcolumn}
\usepackage{bm}

\begin{document}


\title{Giant Angular Dependent Nernst Effect in the Q1D Organic conductor (TMTSF)$_2$PF$_6$}

\author{Weida Wu}
    \altaffiliation[Current address: ]
    {Department of Physics, University of Texas,  Austin, Texas 78712}
\author{N.P. Ong}
\author{P.M. Chaikin}
\affiliation{
Department of physics, Princeton University, Princeton,
New Jersey, 08544}

\date{\today}

\def\tmt{(TMTSF)$_2$X}
\def\tmtx{(TMTSF)$_2$X, (X=PF$_6$, ClO$_4$ ...) }
\def\pf6{(TMTSF)$_2$PF$_6$}
\def\clo4{(TMTSF)$_2$ClO$_4$}
\def\RA{\Rightarrow}
\def\ra{\rightarrow}
\def\lag{\mathcal{L}}
\def\prl{{\it Phys. Rev. Lett.}}
\def\prb{{\it Phys. Rev. B}}
\def\mvk{$\mu V/K$}
\def\ie{{\it i.e.}}
\def\eg{{\it e.g.}}
\def\va{{\bf a}}
\def\vb{{\bf b}}
\def\vc{{\bf c}}
\def\vv{{\bf v}}
\def\vq{{\bf Q}}
\def\etal{{\it et al.}}
\def\ds{\displaystyle}
\def\t1{$T_1^{-1}$}
\def\se{$^{77}$Se}
\def\rzz{R$_{zz}$}
\def\nzx{N$_{zx}$}
\def\ef{\varepsilon _f}
\def\dfermi{\left(-\ppep{f^0}\right)}
\def\oh{\frac{1}{2}}
\def\ddt{\frac{d}{dt}}
\def\ppe{\frac{\partial}{\partial \varepsilon}}
\def\ppkl{\frac{\partial}{\partial k_{_l}}}
\def\dtt{\dot{\theta}}
\def\mt{\stackrel{\leftrightarrow}{\bf T}}
\def\invm{({\mathcal M}^{-1})}
\def\ie{{\it i.e.}}
\def\eg{{\it e.g.}}
\def\X{{\bf X}}
\def\bu{{\bf U}}
\def\ifs{\int_{_{FS}}}
\def\tfp{\frac{\tau}{4\pi^3}}
\def\ppkx{\frac{\partial}{\partial k_x}}
\def\ksl{\ds 1+\left(\pp{k_x}{k_y}\right)^2+
\left(\pp{k_x}{k_z}\right)^2}
\def\be{\begin{equation}}
\def\ee{\end{equation}}
\def\dg{$^{\circ}$\ }
\def\ls{\stackrel{<}{\sim}}
\def\gs{\stackrel{>}{\sim}}
\def\htt{\tilde{t}}
\def\va{\textbf{a}}
\def\vb{\textbf{b}}
\def\vc{\textbf{c}}
\def\obt{\omega^{\theta}_b}
\def\oct{\omega^{\theta}_c}
\def\obtt{\omega^{\theta}_b\tau}
\def\octt{\omega^{\theta}_c\tau}
\def\dg{$^{\circ}$}
\def\go{\gamma\frac{\obt}{\oct}}
\def\bo{\beta\frac{\oct}{\obt}}
\def\jgo{J_1\left(\gamma\frac{\obt}{\oct}\right)}
\def\jbo{J_1\left(\beta\frac{\oct}{\obt}\right)}

\renewcommand{\vec}[1]{{\bf #1}}

\newcommand{\pp}[2]{\frac{\partial #1}{\partial #2}}
\newcommand{\ppp}[1]{\frac{\partial}{\partial #1}}
\newcommand{\ppep}[1]{\frac{\partial #1}{\partial \varepsilon}}
\newcommand{\dd}[2]{\frac{d #1}{d #2}}
\newcommand{\ts}[1]{{\mathcal #1}}
\newcommand{\grad}[1]{\nabla #1}
\newcommand{\bk}[1]{\left<#1\right>}
\newcommand{\abs}[1]{\left|#1\right|}
\newcommand{\mr}[1]{\frac{1}{1+(#1)^2}}
\newcommand{\hall}[1]{\frac{#1}{1+(#1)^2}}
\newcommand{\pemr}[1]{\frac{1}{\sqrt{2}t_a}\frac{2(#1)^2}{[1+(#1)^2]^2}}
\newcommand{\pemrr}[1]{\frac{2(#1)^2}{[1+(#1)^2]^2}}
\newcommand{\pehall}[1]{\frac{-1}{\sqrt{2}t_a}\frac{(#1)[1-(#1)^2]}
{[1+(#1)^2]^2}}
\newcommand{\pehh}[1]{\frac{(#1)[5+7(#1)^2]}{[1+(#1)^2]^2}}

\newcommand{\figeps}[3]
    {
    \begin{figure}[htbp]
    \centerline{\includegraphics[width=#2truein]{#1.eps}}
    \caption{\label{#1}#3}
    \end{figure}}

\newcommand{\figepss}[3]
    {
    \begin{figure*}[htbp]
    \centerline{\includegraphics[width=#2truein]{#1.eps}}
    \caption{\label{#1}#3}
    \end{figure*}}

\voffset=1cm
\begin{abstract}
We present a detailed study of the Nernst effect N$_{zx}$ in
(TMTSF)$_2$PF$_6$ as a function of temperature, magnetic field
magnitude and direction and pressure. As previously reported there
is a large resonant-like structure as the magnetic field is rotated
through crystallographic directions, the Lebed Magic Angles. These
Nernst effect resonances strongly suggest that the transport of the
system is effectively ``coherent'' only in crystallographic planes
along or close to the applied field direction. We also present
analytical and numerical calculations of the conductivity and
thermoelectric tensors for (TMTSF)$_2$PF$_6$, based on a Boltzmann
transport model within the semi-classical approximation. The
Boltzmann transport calculation fails to describe the experiment
data. We suggest that the answer may lie in field induced decoupling
of the strongly correlated chains.
\end{abstract}

\maketitle

\section{Introduction}
\pf6 is a quasi-one-dimensional electronic system, which displays various
ground states ranging from triplet superconductor\cite{Lee1997, Lee2002K,
Lee2002U} to spin-density wave(SDW) insulator, depending on pressure,
temperature and magnetic field(for a review, please see
Ref.\onlinecite{Ishiguro1997}). \pf6 consists of
plate-like TMTSF molecules which stack with strong wavefunction
overlap in chains. The intrachain bandwidth is $\sim$ 1eV while the
interchain couplings give anisotropic bandwidths of 0.1eV and 0.003ev
in the approximately orthogonal directions. In the `metallic' phase
under moderate magnetic field, a fascinating phenomenon, the so-called
Lebed Magic Angle Effect (MAE) was discovered\cite{Osada1991, Naughton1991,
Kang1992MA} after Lebed's initial prediction\cite{Lebed1986, Lebed1989}.
The first manifestations of these MAEs were sharp
resistance dips when the magnetic field was aligned at inter-chain
directions in real space (lattice vectors\cite{Strong1994}).
In reciprocal space a field along the magic angles induces
electron motion along commensurate {\bf k} space orbits\cite{Osada1992}.
Despite many theoretical efforts to describe the magic angle
effects\cite{Osada1992, Osada1998, Maki1992, Yakovenko1992, Chaikin1992,
Strong1994, Lebed1996, Ong2004}, there is as yet no
satisfactory explanation.  Most of the theories
focus on the semiclassical motion of electron on the open Fermi surface
derived from single particle band structure.

Recently, a giant Nernst
effect was discovered in \pf6. As the magnetic field was rotated toward a
magic angle the Nernst signal increased then
decreased toward zero, changed sign at the magic angle and continued in an
inverse manner. The result is
a sharp resonant-like structure.
\cite{Wu2003} The magnitude of the Nernst signal at 1K is at least 3 orders
larger than what we expected from simple (Drude) estimates. The sign change
of the Nernst effect at the magic angles strongly suggests that the transport
involved in the Nernst effect is effectively 2-dimensional at these commensurate
angles. Both the sign change at the magic angles and magnitude of the signal
are not yet explained, but the effect appears generic for these materials.
The giant resonant Nernst voltage has recently been observed in the sister
compound \clo4.\cite{Choi2005}
Present phenomenological models for the sign change involve field induced
inter-plane decoupling.\cite{Strong1994} Although there is some experimental
evidence for this decoupling, there are not yet theoretical models which
rigorously demonstrate this phenomenon. Giant Nernst signals have also
been seen in high transition temperature superconducting (HT$_c$) cuprates
where a model invoking superconducting
vortices and 2D superconducting phase coherence has been
successful.\cite{Wang2001, Wang2002p, Wang2002s}.  A similar model has been
proposed for \pf6 \cite{Ong2004}.  However, to apply this idea in \pf6 is quite
controversial. On one hand it
naturally explains the large Nernst signal with undetectable
thermopower signal, predicts a particular sign of the Nernst effect
confirmed by experiments and qualitatively explains some aspects of
experiments.  On the other hand, it predicts a large superconducting
fluctuation region in the phase diagram, which is absent in other
measurements.  Most of the superconducting properties in \pf6
have been understood within a conventional mean field BCS picture.
We will explore the possibility of this vortex Nernst effect with
more experimental detail in a subsequent paper.

What sorely hampers progress in understanding these unusual magic angle
phenomena are the lack of measurements other than charge transport.
The Field Induced SDW (FISDW) and MAE have been observed primarily
in the charge channel by transport measurements. The FISDW has been
more thoroughly explored with magnetization, magnetocaloric effect and
spin relaxation studies. Magnetic torque measurements on
(TMTSF)$_2$ClO$_4$ suggest there is a thermodynamic component to the MAE
\cite{Naughton1991}. While a thermodynamic probe is
an obvious choice for establishing the presence of unknown phases
or fluctuations in \pf6, the high pressure environment makes
a measurement of specific heat or dc magnetic susceptibility impractical.
Recently, the \se\ NMR spin-lattice relaxation rate
measurements\cite{Wu2005} at different magnetic field orientation show
no evidence for either a spin gap or a single particle gap.
Furthermore, there is no evidence for an enhancement of the FISDW
transition temperature. This strongly suggests that neither FISDW ordering
nor fluctuations are likely to be responsible for the MAE.
The dramatic contrast between the charge channel and the spin channel
at MAs suggests that spin and charge degrees
of freedom may be decoupled.\cite{Vescoli1998, Lorenz2002}
The thermodynamic and suggested coherent-incoherent transitions
would therefore be the result of interaction and correlation
effects due to subtle changes in the electronic wavefunctions and
density wave susceptibilities.

Before speculating further on exotic mechanisms for the giant Nernst
resonances and other MAEs in transport it is necessary
to see what conventional transport theory will yield.
Although Boltzmann transport calculations as a function of magnetic field
magnitude and direction have been performed for resistance\cite{Osada1992}
there has been no such study for the thermoelectric transport
coefficients. Such calculations are one of the main contributions
of this paper.

We divide our presentation into two sections.  The first
section focuses on the Nernst experiments.  We present a detailed study of
the Nernst effect \nzx in \pf6 at various pressures, magnetic fields and
temperatures. The second part presents both numerical and analytic
calculations of Boltzmann transport in the
relaxation time approximation with realistic band parameters . We then
compare the calculations with our experiment data.

\section{Nernst measurement}
\subsection{Method}
Figure \ref{setup} shows the experimental setup for the Nernst
effect \nzx measurement with the temperature gradient along the
\va-axis and voltage measured along \vc-axis. 3 pairs of Au wires
were attached to the opposite \va\vb-planes of the sample by silver
paint for both resistance measurements(the end pairs) and the Nernst
measurements(the middle pair). The Au wires were attached to the
alloy wires (Phospher Bronze) fed through the pressure cell base.
Here we used low thermal conductivity alloy wires instead of Cu
wires to minimize the possible transverse temperature gradients. A
miniature heater was placed on top of the sample to establish a
small temperature gradient along the \va-axis.  Two thin film RuO
thermometers were used to measure the temperature difference.  The
thermoelectric voltage is measured by a Keithley 182 Nanovoltmeter.
The heater was turned on and off for a few cycles for signal
averaging.\cite{Yu_thesis1990} A linear-fit-extrapolation method was
used to accommodate the slow drift of the baseline
signal.\cite{Wu_thesis2004} The resistance was measured by a
conventional 4-probe low frequency lock-in technique. The
magneto-resistance \rzz and  Nernst signals \nzx were measured
simultaneously.

\figeps{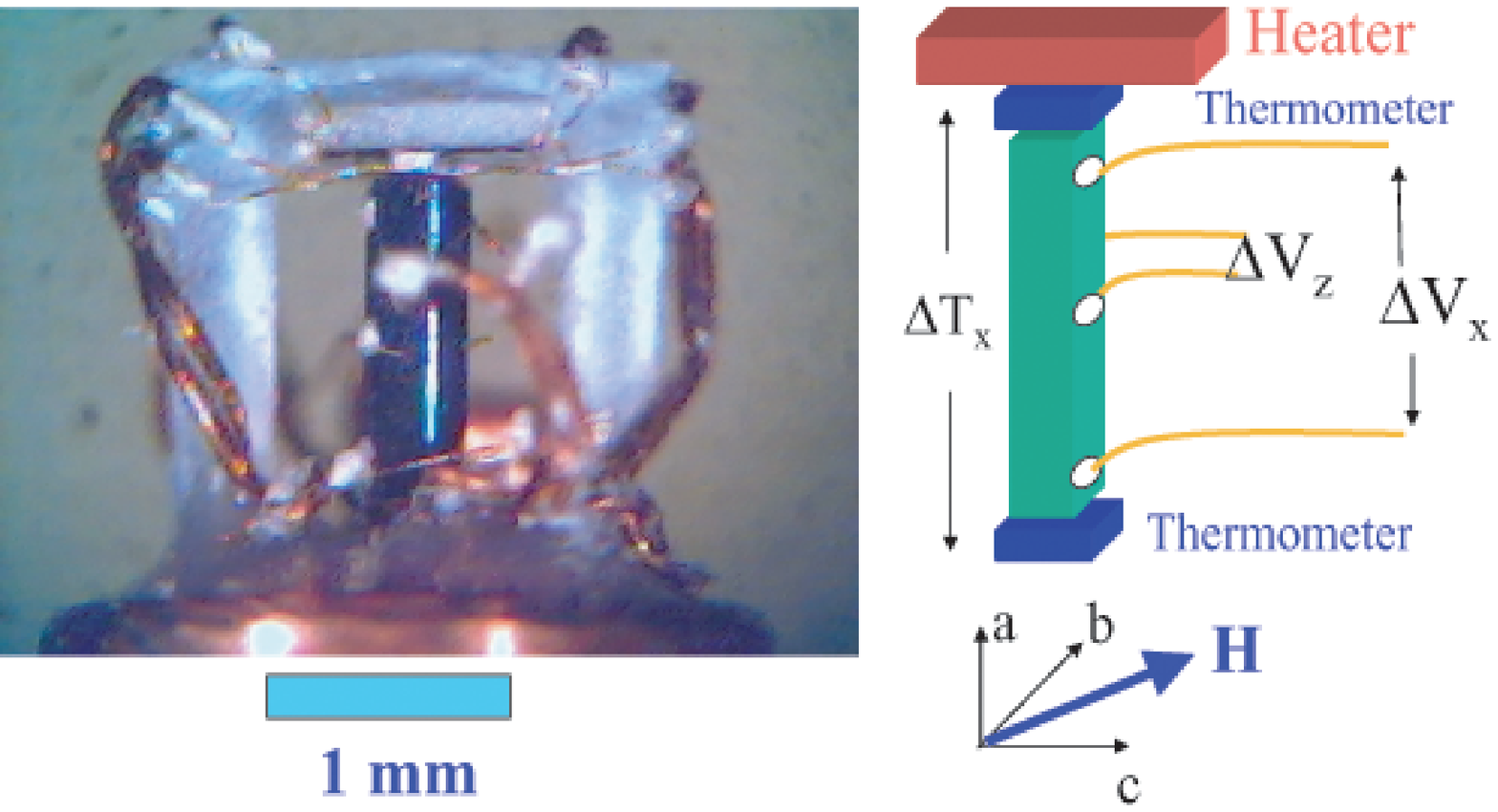}{3}{The measurement setup of Nernst effect \nzx: Three
pairs of Au wires were attached to the sample along the {\bf a} axis
on the opposite sides of the \va\vb-planes of the sample. Two RuO
thin film resistance thermometers were placed next to both ends of
the sample to measure the temperature gradient generated by a
miniature heater on the top. The middle pair of leads was used for
Nernst voltage $\Delta V_z$ pickup. The other two pairs of leads
were used for 4-probe interplane (\vc-axis) resistance measurement.
The magneto-resistance, $R_{zz}$, was measured simultaneously with
the Nernst effect.}

Fig. \ref{NstRzzQT1KB6T} shows a typical angle dependence of the
Nernst signal \nzx in \pf6,  obtained at 1 K, 6 Tesla and 13 kbar.
The magnetic field was rotated from -40$^{\circ}$ to 50$^{\circ}$
with respect to the \vc*-axis.(See Ref.\onlinecite{Wu2003}
for definitions of \vc*, \vc' etc.)
The maximum Nernst signal is about 100 $\mu$V/K, found
at approximately 3$^{\circ}\sim 4^{\circ}$ off \vc'
($\theta_{c'}=7^\circ$). The Nernst coefficient is of the order of
10 $\mu$V/K$\cdot$T. As far as we know, this value is much larger than the
Nernst effect observed in any other metal. The angular dependence of the
Nernst signal agrees well with our previous thermoelectric measurement in
a different geometry.\cite{Wu2003} To study the temperature, field and
pressure dependence of the giant Nernst effect, we fixed the magnetic
field orientation at 3$^{\circ}$ off a magic angle(\vc' or -1L)
.  First, let's discuss the sign of
the Nernst effect in \pf6.  This is very important for
the vortex Nernst model.

\figeps{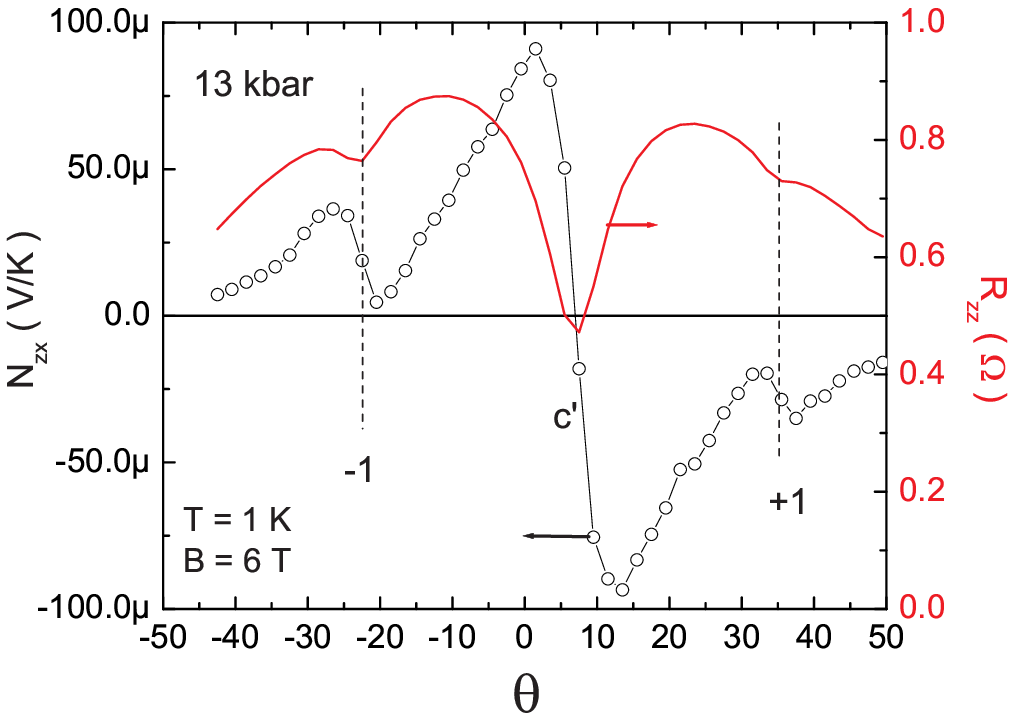}{3}{Angle dependence of the Nernst signal $N_{zx}$
and \vc-axis resistance $R_{zz}$ were measured simultaneous
at 1 K, 6 Tesla and 13 kbar. The thin line data are $R_{zz}(\theta)$.
The open circles ``$\circ$'' data are the Nernst data $N_{zx}(\theta)$.
The solid line is a guide to eyes.  The Nernst resonances are
well aligned with the magic angles marked by the resistance dips.
Here $\pm 1$ correspond to inter-chain directions \vc'$\pm$\vb'.}

\subsection{The sign of the Nernst effect}
In the vortex liquid phase of a type II superconductors vortices
flow down the temperature gradient,\vec{v}$\parallel(-\nabla T)$ and
generate an electric field $\vec{E}=\vec{B}\times\vec{v}$ transverse
to the temperature gradient $-\nabla T$ according to Josephson
relation\cite{Josephson1962, Josephson1965}. Therefore, the sign of
the vortex Nernst effect is fixed by $\nabla T\times\vec{B}$. In
general, the Nernst effect of an electronic system can have either
sign depending on details of the band structure. To determine the
sign of the Nernst signal, we noted the orientation of the sample
and leads, and placed an alignment mark on the base/feedthru of the
pressure cell and on the cell body. We assume the alignment mark
doesn't change much on pressurization. To get the Nernst sign
correct we need to know the orientation to better than $90^{\circ}$.
We see the magic angles where we expect them to be to $\sim
15^{\circ}$ We observed no orientation variation when the pressure
was increased in the same pressure cell. Our measurements show that
the sign of the Nernst effect is \emph{consistent} with the vortex
Nernst model (but certainly does not prove it).

\subsection{Temperature dependence}
\figeps{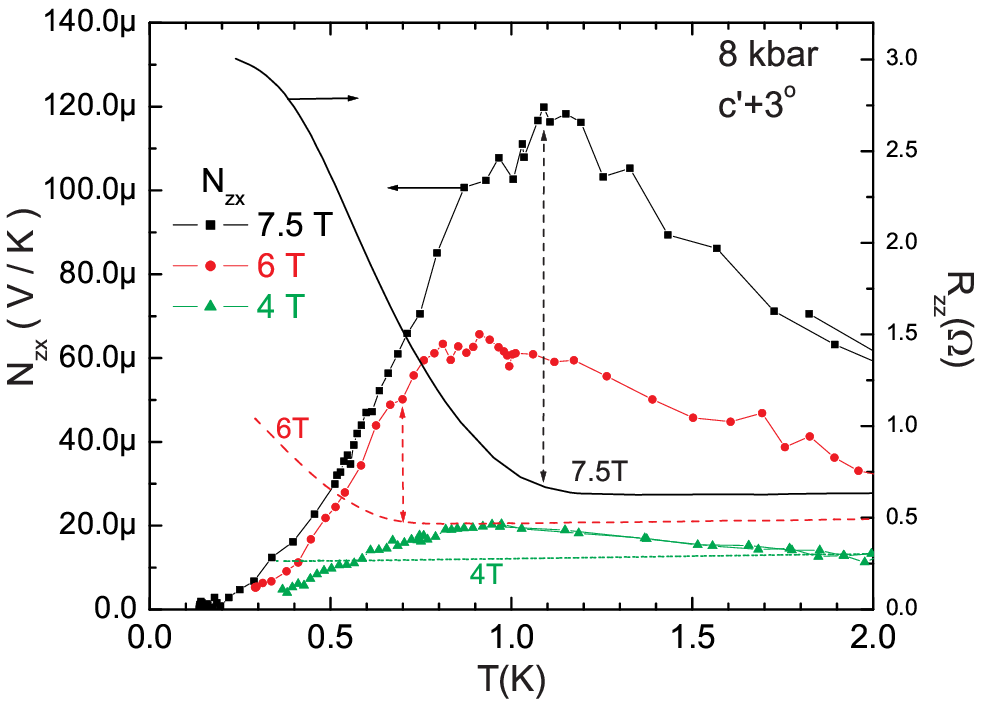}{3}{Simultaneous measured temperature dependences
of the \vc-axis resistance \rzz\ (thin lines) and the Nernst signal
\nzx (filled symbols) at 8 kbar for 3$^{\circ}$ off \vc'. The Nernst
signals rise to a maximum around 1 K, then fall exponentially and
are unmeasurable below $\sim$200 mK. The decrease of the Nernst
signal is correlated with the upturn of the resistance as T
decreases. The upturn of R$_{zz}$ indicates the Metal-FISDW phase
transition.}

Fig.\ref{NRT8kbar} shows the temperature dependence of the Nernst
effect \nzx\ at \vc' at different field values: 4, 6 and 7.5 Tesla
for 8 kbar pressure.  The Nernst signals rise gradually to a maximum
around 1 K as the temperature decreases, then fall off roughly
exponentially at lower temperature and are unmeasurable below $\sim$
200 mK. Clearly, the Nernst signal is non-linear with magnetic field
below 2 K. The temperature dependence of the Nernst signal at 7.5
Tesla agrees with the previous Nernst data from the thermopower
measurement.\cite{Wu2003} At high magnetic fields, \pf6 enters the
FISDW insulting phase at a critical temperature T$_c$(H) determined
by the sharp rise of resistance \rzz\ measured simultaneously. In
the ``metallic'' phase, the Nernst effect at \vc' and $\pm$1 Lebed
angles have similar temperature dependence, except the magnitude is
much larger for \vc'. However,  the presence of the FISDW phase
seems to suppress the Nernst signal at \vc'. As shown in
Fig.\ref{NRT8kbar} for 7.5 Tesla data, a sudden decrease of the
Nernst voltage happens at the onset of FISDW transition.
\figeps{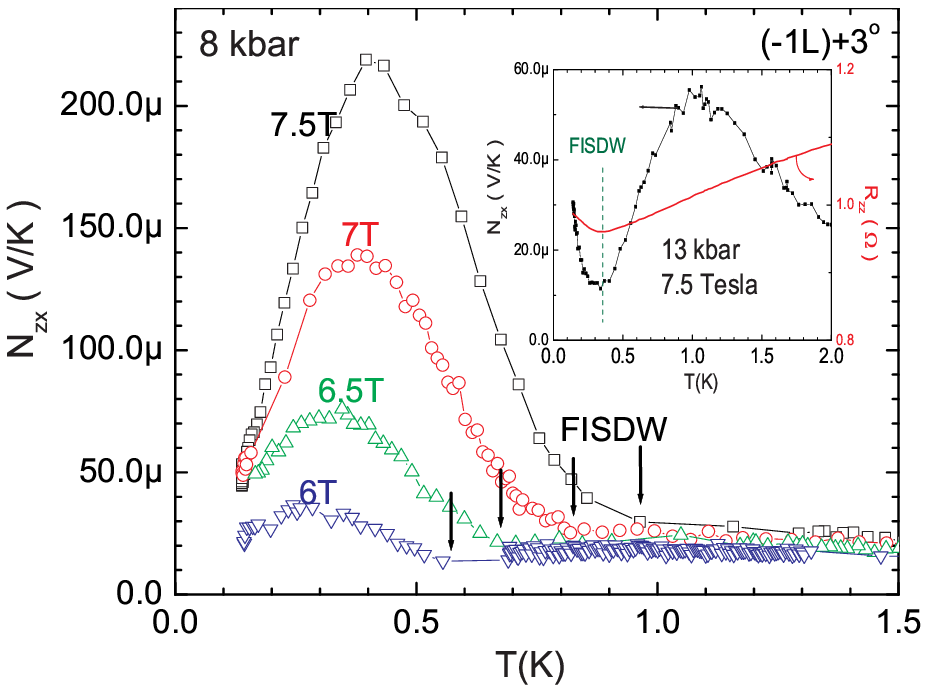}{3}{Main panel: Temperature dependence of the
Nernst signal \nzx measured at 3\dg away from -1 Lebed angle at 8
kbar. The magnetic fields are: 7.5, 7, 6.5 and 6 Tesla respectively.
The arrows mark the FISDW transition temperature obtained from
resistance measurements R$_{zz}$(T) (not shown for clarity) at
various fields. Insert: Temperature dependence of the Nernst signal
\nzx and resistance \rzz of -1 Lebed angle at 13 kbar for 7.5 Tesla
magnetic field. The dash line marks the FISDW transition ($\sim$
350mK).  An enhancement of Nernst voltage was found in the FISDW
phase.}

Interestingly, the presence of the FISDW phase affects the Nernst signal
at -1 Lebed angle differently. As shown in Fig.\ref{NTm1L8kbar}, at 8 kbar
pressure the FISDW onsets (from R$_{zz}$) coincide the onsets of
a large increase of Nernst voltage at -1 Lebed angle at various magnetic fields.
At still lower temperature the voltage at the -1 Lebed angle
reaches a peak around 300 $\sim$ 400 mK then decreases quickly.
The peak value is as large as $\sim$ 220 $\mu$V/K at 7.5 Tesla.
This is further confirmed by the angle dependence of the Nernst effect
at base temperature (150 mK), where there is large Nernst resonances at
$\pm$1 Lebed angles while there is none at \vc'. This behavior is
consistent with our previous measurements.\cite{Wu2003}
The effect of FISDW on -1 Lebed angles is further confirmed by measurements
at higher pressures,
where FISDW transition temperature T$_c$(H) vary accordingly.
For example, at 13 kbar the FISDW transition temperature at 7.5 Tesla
is suppressed down to $\sim$ 350 mK, the enhancement of
the Nernst effect at -1 Lebed angle follows the FISDW transition
accordingly as shown in the insert of Fig.\ref{NTm1L8kbar}.

The suppression of the Nernst signal at \vc' by the FISDW is not
understood at this moment.  It is probably due to the competition between
the FISDW phase and ``metallic' phase.
This difference seems to suggest the magic angle \vc' is different
from -1 Lebed magic angle in a subtle way. In this paper, we limit
our discussion within the ``normal'' state where the MAE is pronounced.
We note that a full understanding of the MAE should cover the FISDW
phase, where the MAE is more complicated than in the metallic phase.

\subsection{Field dependence}
\figeps{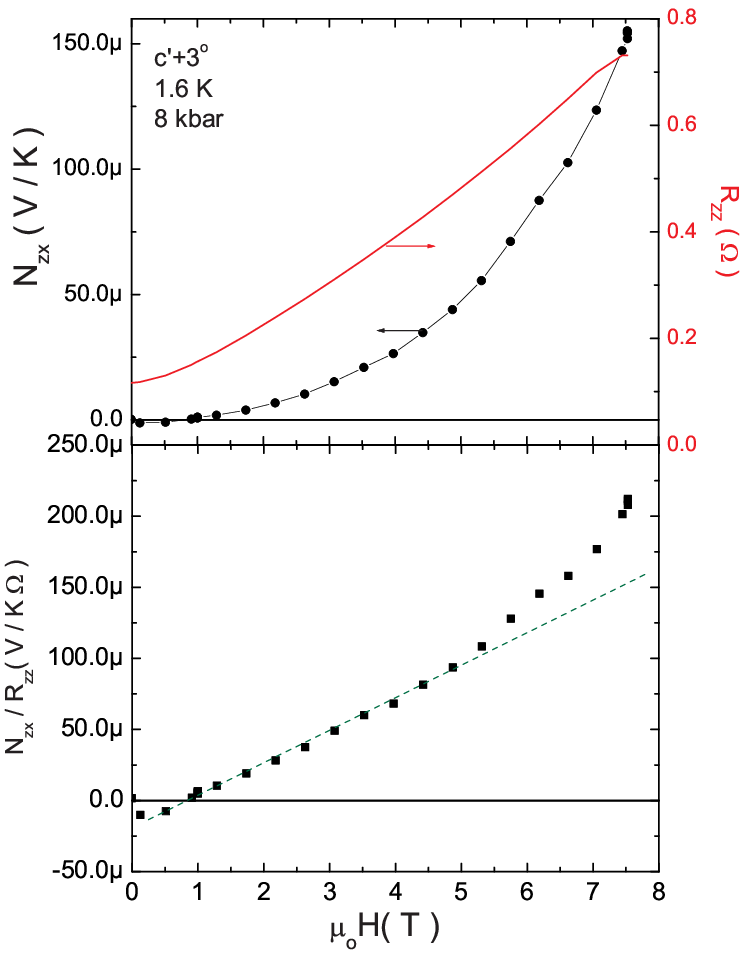}{3}{Upper panel: Magnetic field dependence of \nzx\
and \rzz\ measured simultaneously at 8 kbar, 1.6 K for \vc'. Clearly
\nzx\ is non-linear with magnetic field just as observed
in previous measurement\cite{Wu2003}. \\
Lower panel: Ratio of \nzx\ and \rzz\ derived
from the upper panel. As discussed in text,
$\ds\frac{N_{zx}}{R_{zz}}\propto\alpha_{zx}$.
The dash line is a guide to eyes.  It is clear that at low field
$\alpha_{zx}$ is linear with field. }

As seen in the temperature dependence of \nzx\ at different magnetic
fields, the Nernst effect in \pf6 is very non-linear with magnetic
field. In the upper panel of Fig.\ref{NAH8kbar}, we show
simultaneous measurements of magneto-resistance \rzz\ and the Nernst
effect \nzx\ vs. magnetic field at 1.6 K, 8 kbar and 3\dg\ off \vc'.
It is clear that the Nernst signal has a super-linear field
dependence. An obvious nonlinear effect is the large
magnetoresistance of \pf6. In transport theory, the thermopower
tensor $\ts{S}$ is the product of the resistivity tensor $\ts{\rho}$
and the thermoelectric tensor $\ts{\alpha}$.

\be \ts{S}=\ts{\rho}\cdot\ts{\alpha} \label{e:S} \ee

\figeps{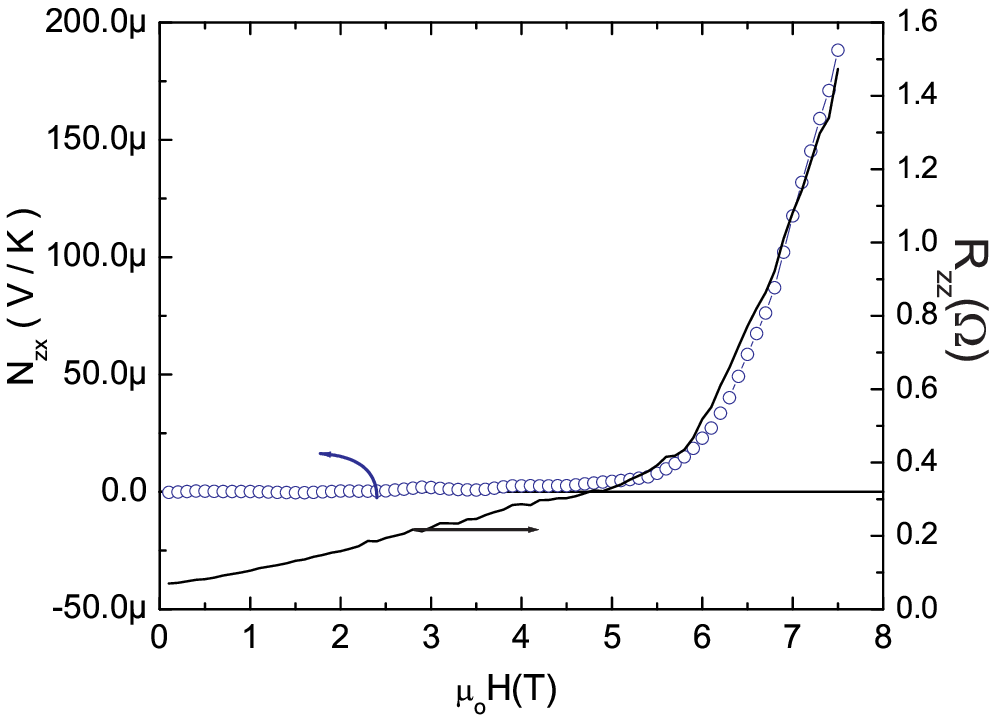}{3}{Magnetic field dependence of \nzx\ and \rzz\
measured simultaneously at 8 kbar and 375 mK for field at 3\dg\ away
from -1 Lebed angle. The upturn of the \rzz\ around 5.5 Tesla
defines the threshold field of the FISDW transition. \nzx\ is highly
non-linear and enhanced greatly in FISDW phase.}

Therefore, $\ts{S}_{zx}=\rho_{zx}\alpha_{xx}+\rho_{zy}\alpha_{yx}
+\rho_{zz}\alpha_{zx}\approx\rho_{zz}\alpha_{zx}$. Here we ignore
the first two terms since the Hall effects are negligibly small in
the metallic phase for \pf6.  To obtain $\alpha$, we took
the ratio of the Nernst signal \nzx\ and the resistance \rzz\
at the same field to obtain field dependence of
$\ds \alpha_{zx}\propto\frac{N_{zx}}{R_{zz}}$ according to Eq.\ref{e:S}.
The prefactor depends on the sample geometry. The result is
shown in the lower panel of Fig.\ref{NAH8kbar}.  Clearly
$\ds\alpha_{zx}$ is approximately linear with
magnetic field below 5 Tesla.  Therefore, the non-linearity
of the Nernst signal \nzx\ in \pf6 mainly comes from
the large magneto-resistance.
The linear field dependence for field along \vc' and -1 Lebed angle
suggests that $\alpha_{zx}$ is probably a more fundamental
quantity in the thermoelectric effect of \pf6.

Fig.\ref{NRH8kbar1L} shows the field dependence of $N_{zx}$ and
$R_{zz}$ for field along 3\dg\ off the -1 Lebed angle at 375 mK. We
can see that when the \pf6 goes into the FISDW, the resistance rises
up sharply around 5.5 Tesla due to the presence of the FISDW gap.
The Nernst signal also shows a sharp upturn around 5.5 T and rises
up dramatically.  This agrees with the observation of the
enhancement of the Nernst signal at -1 Lebed angle in the
temperature dependence (Fig. \ref{NTm1L8kbar}).

\subsection{The effect of pressure}
\figeps{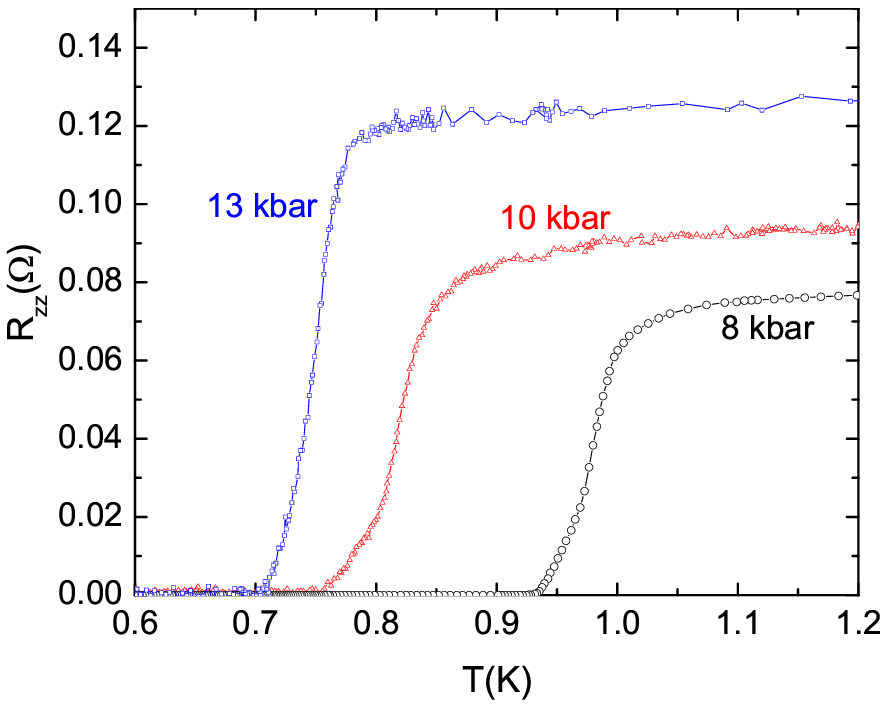}{3}{Zero field temperature dependence of R$_{zz}$
at 8, 10 and 13 kbar. The T$_{c0}$'s are 1.02 K, 0.87 K and 0.77 K
respectively.}

The ground state properties of the Bechgaard salts are strongly
affected by hydrostatic pressure. The temperature, pressure and
magnetic field (T-P-H) 3D phase diagram (Fig.1 in
Ref.\onlinecite{Kang1993}) summarizes the effects of pressure on
various phase transitions. For example, the threshold field of the
FISDW phase progressively increases as the pressure gets higher. The
superconducting transition temperature T$_c$ is also slowly
suppressed by increasing pressure. Fig.\ref{TcCompare} shows the
zero field temperature dependence of resistance \rzz\ at 8 kbar, 10
kbar and 13 kbar respectively. The superconducting transition
temperature decreases slowly with increasing pressure. Using the
onset definition(90\% of the normal state value), we found that the
superconducting transition temperatures T$_{c0}$ are 1.02 K, 0.87 K
and 0.77 K respectively. \figeps{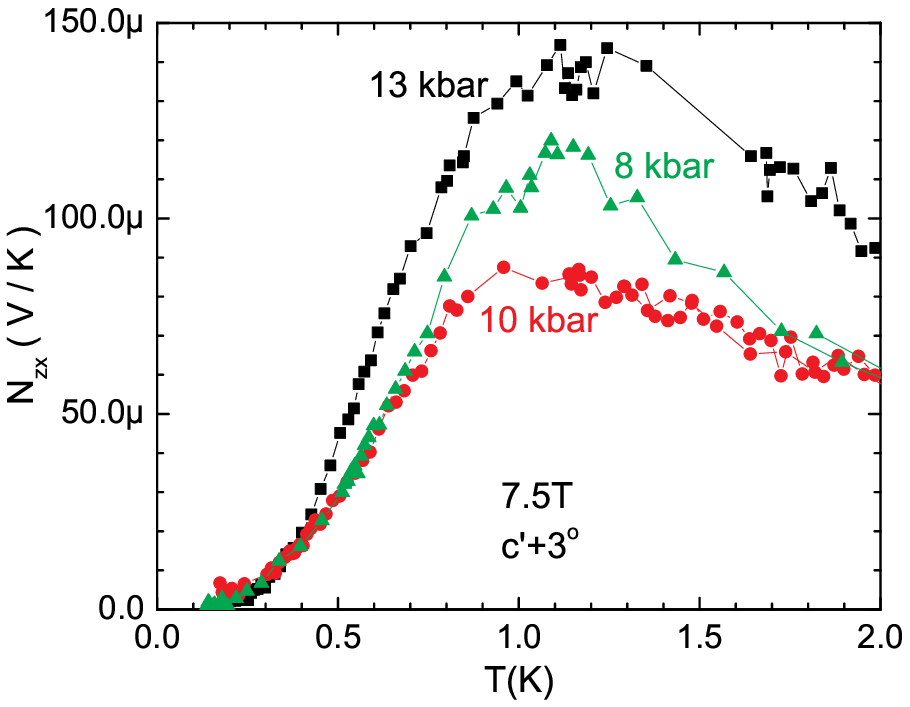}{3}{Temperature
dependence of the Nernst signal \nzx\ at 8 kbar , 10 kbar and 13
kbar at 7.5 Tesla.  Qualitatively no change is observed for various
pressures.} In Fig. \ref{NTPcom7p5T} we show the comparison of the
temperature dependence of the normalized Nernst signal at
3$^{\circ}$ off \vc' between 8 kbar, 10 kbar and 13 kbar.
Qualitatively, the temperature dependence of the Nernst signal is
pressure insensitive. \figeps{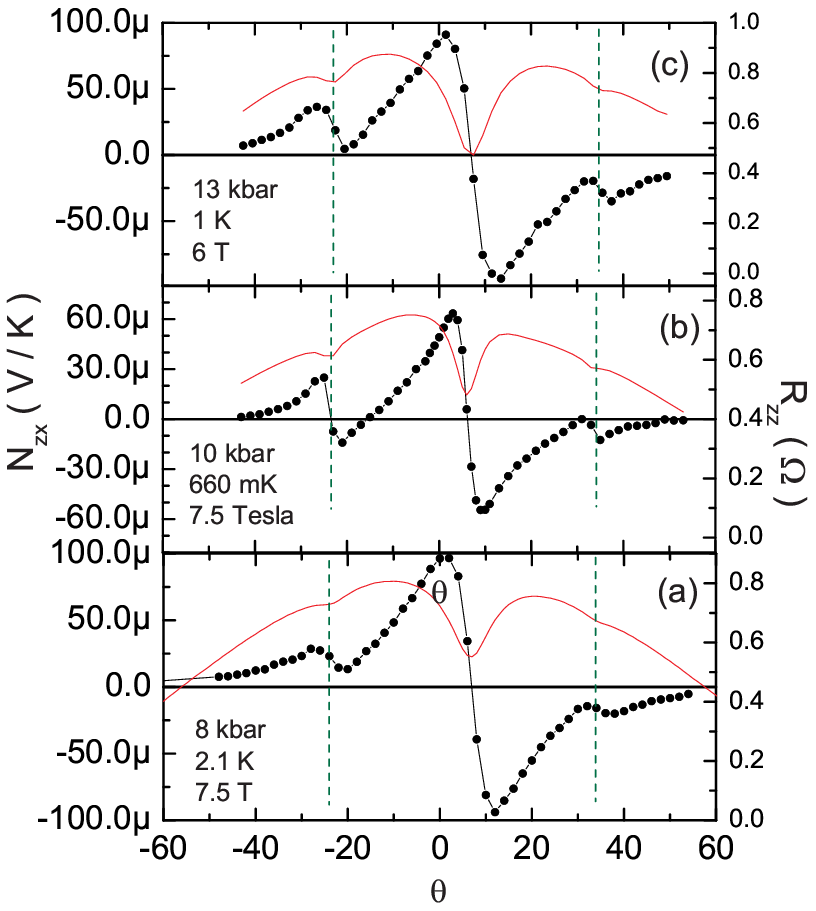}{3}{Angle dependence of \nzx
(``$\bullet$'') and \rzz (thin line) at (a) 8 kbar, 2.1K and 7.5
Tesla; (b) 10 kbar, 660mK and 7.5 Tesla; (c) 13 kbar, 1K and 6
Tesla. Qualitatively, no change is observed. }

The angular dependence doesn't change significantly either as we vary the
pressure. Fig. \ref{NRQPcom} shows the angular dependence of \nzx\
and \rzz\ at 8, 10 and 13 kbar respectively.
Note that the temperature and magnetic fields of these data
are not identical.   However, qualitatively they are all the same.
Note that all data shown here are in the \emph{metallic} phase.

\section{Boltzmann calculation of Nernst effect}
To gain some elementary intuition about transport processes
it is often instructive to look first at a generalized
Drude approximation, by which we mean a classical gas of charged particle
in lowest order response to an applied set of driving fields.
\subsection{Drude Transport}

\figepss{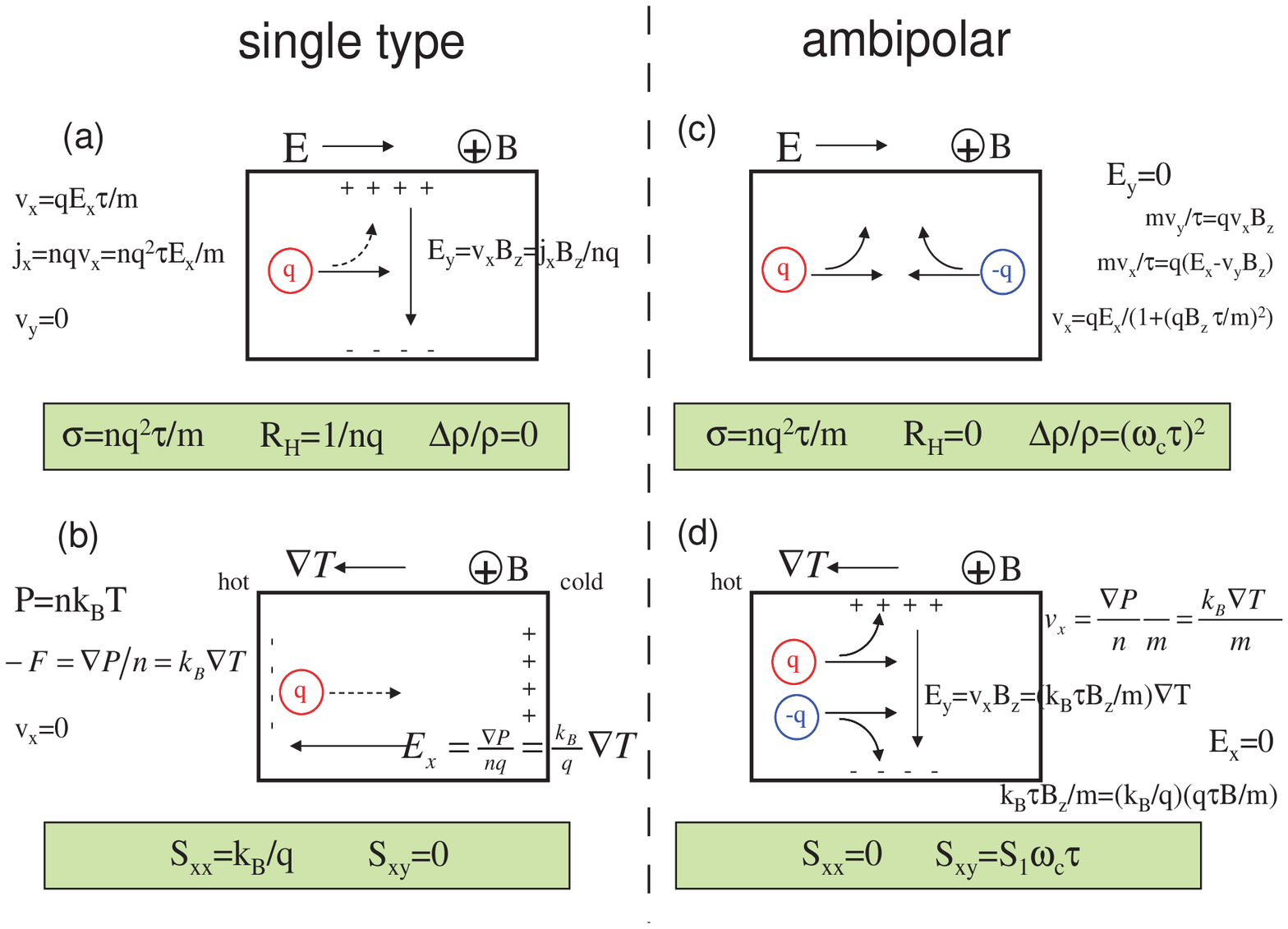}{6}{Generalized Drude model schematic. Magnetic field
perpendicular to the plane. (a), (c) E field drive, (b), (d) $\nabla
T$ drive, (a),(b), one carrier, (c),(d) two oppositely charged
carriers.}

In Fig.\ref{drude} we show a cartoon of the particle motion of such
a charged gas. In a Drude model forces accelerate particles which
then lose momentum in collisions at a rate $1/\tau$. The basic
equation of motion is therefore $m\vec{v}/\tau = \vec{F}$, the
charge per particle is q, the particle density is n, the current
density is simply charge density times velocity $\vec{j}=nq\vec{v}$
and $\vec{j}=\sigma\cdot\vec{E}$. In Fig.\ref{drude} (a), $\vec{F}$
is $q\vec{E}$ and $\ds\vec{v}=\frac{q\tau}{m}\vec{E}$,
$\ds\vec{j}=\frac{nq^2\tau}{m}\cdot\vec{E}$, the conductivity
$\sigma$ is given by $\ds\sigma = \frac{nq^2\tau}{m}$. In the
presence of the magnetic field there is a Lorentz force which
deflects particles in the $y$ direction building up charges on the
upper and lower boundaries. The charges continue accumulating until
the electric field they generate exactly cancels the Lorentz force,
$E_y=v_x\cdot B_z$. In steady state the Hall field
$E_y=\frac{j_x}{nq}\cdot B_z$ completely compensates  the effect of
the magnetic field, the carriers only drift in the x direction and
there is no magnetoresistance. In (b) the drive is a temperature
gradient. In this simple model the gas is ideal and we use the ideal
gas law, $P=nk_BT$. A temperature gradient translates to a pressure
gradient $\nabla P= n k_B \nabla T$, or a force per particle of
$\ds\vec{F}=\frac{\nabla P}{n}$. The charged particles will flow in
the x direction charging the boundaries and creating an opposing
field $E_x$. The current and charging stop when $qE_x=F_x$ or $\ds
E_x=\frac{k_B}{q}\nabla_x T$. This thermoelectric voltage is the
Seebeck effect with coefficient $\ds S_1=\frac{k_B}{q}$. (This was a
big failure of the Drude model. It overestimates the thermopower by
several orders of magnitude. The reason is quite evident today. We
have a degenerate electron gas (DEG) rather than an ideal classical
gas. The effective number of degrees of freedom, or particles that
can transport heat, is reduced by $\ds\sim \frac{k_BT}{E_F}$ so the
$\ds S_{1DEG} \sim \frac{k_BT}{E_F}\frac{k_B}{q}$). In this picture
the electric force cancels the pressure and the particles have no
velocity, $\vec{v} \times \vec{B}=0$ and the Nernst voltage is zero.
For the simplest conductors we therefore expect the thermopower to
be sizeable and the Nernst effect negligible. But it is worth noting
that effectively the same argument would suggest that the
magnetoresistance is negligible.

In Fig.\ref{drude} (c) and (d) we consider the case of two
oppositely charged carriers, which are otherwise identical. With an
electric field along x the two carriers move at $\ds
v_x=\frac{q\tau}{m}E_x$ in opposite directions both contributing to
the electrical current and conductivity which remains $\ds\sigma_0 =
\frac{nq^2\tau}{m}$. Now however, in the presence of $B_z$ both are
deflected in the same direction, there is no charge accumulated on
the boundaries, there is no Hall voltage, and velocities persist in
both directions, $\ds v_y=v_x\frac{q B_z\tau}{m}$, $\ds
v_x=\frac{q\tau}{m}E_x - v_y\frac{q B_z}{m}\tau$, with the solution,
$\ds v_x = \frac{qE_x}{1+(\frac{qB_z\tau}{m})^2}$, $\ds\sigma=
\frac{\sigma_0}{1+(\omega_c \tau)^2}$ where $\ds\omega_c =
\frac{|q|B_z}{m}$. There is now magnetoresistance,
$\ds\frac{\Delta\rho}{\rho}=(\omega_c \tau)^2$. In (d) the drive is
a temperature gradient again producing a pressure gradient. Both
types of particles move down the pressure gradient with velocity,
$\ds v_x = \frac{k_B \nabla_x T \tau}{m}$ there is no charge
accumulation, and no field generated along x so the Seebeck
coefficient is zero. In the presence of a magnetic field, the
particles with the same velocity but opposite sign are separated,
charges accumulate on the upper and lower boundaries until the
electric field compensates the Lorentz force, $\ds E_y = v_x B_z =
\frac{k_B \nabla_x T \tau B_z}{m}$. The result is a Nernst voltage
with coefficient $\ds S_{yx}=\frac{k_B\tau}{m}
=\frac{k_B}{|q|}\frac{|q|B_z\tau}{m} =S_1 \omega_c \tau$. With the
degenerate electron gas correction we should then expect $\ds S_{xy
DEG} \sim \frac{k_B}{|q|}\frac{k_B T}{E_F}(\omega_c \tau)$.

\begin{tabular}{|c|c|c|c|c|c|}
   \hline
   & $\sigma$ & Hall  & MR
   & Seebeck & Nernst \\ \hline
  one carrier & $\ds\frac{nq^2\tau}{m}$ & $\ds\frac{1}{nq}$
  & 0 & $\ds\frac{k_B}{q}\frac{k_B T}{E_F}$ & 0 \\ \hline
  ambipolar & $\ds\frac{nq^2\tau}{m}$ & 0 & $(\omega_c \tau)^2$ & 0 &
  $\ds\frac{k_B}{|q|}\frac{k_B T}{E_F}\omega_c \tau$
  \\ \hline
\end{tabular}

The extension of these results to field orientation and different
ratios of the densities of the oppositely charged particles is
straightforward. With $\ds S_1 =\frac{k_B}{|q|}\frac{k_B T}{E_F}$,
$\ds a=\frac{n^{+} -n^{-}}{n^{+} +n^{-}}$, $\ds
f(a)=a\frac{1+(\omega_c \tau)^2}{1+a^2(\omega_c \tau)^2}$, $\ds
g(a)=\frac{1-a^2}{1+a^2(\omega_c \tau)^2}$, we find $S_1 (a) = S_1
f(a)$, $S_{xy}(a)=S_1 (\omega_c \tau)g(a)$, $\ds R_H =
\frac{1}{n|q|}f(a)$, $\ds\frac{\Delta\rho}{\rho} = (\omega_c \tau)^2
g(a)$. The Hall and Nernst voltages vary as $\vec{B} \times \vec{E}$
and $\vec{B}\times \nabla T$ respectively. $f(a)$ and $g(a)$ are
plotted in Fig.\ref{Drude_Gen}. Putting in realistic parameters, we
will obtain the Nernst effect in Drude model is order of 10 nV/K,
linear with magnetic field.  Drude picture only predicts that the
Nernst effect has a simple $\sin\theta$ dependence of magnetic field
orientation and linear with magnetic field and temperature.
\figeps{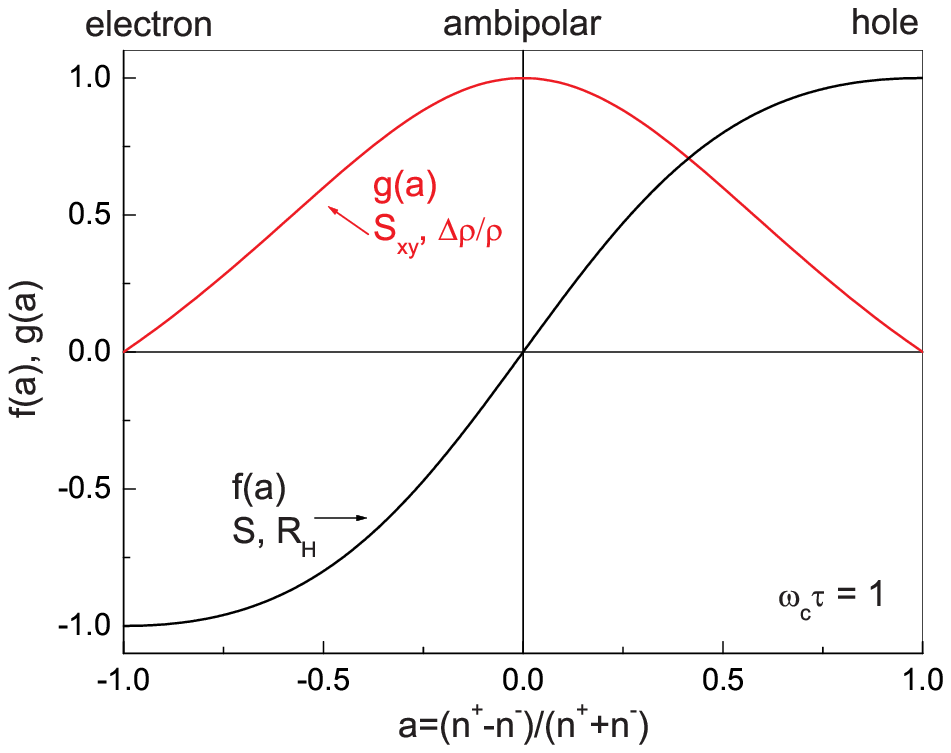}{3.3}{Drude picture in general. Here $\ds
a=\frac{n^{+} -n^{-}}{n^{+} +n^{-}}$ is the net charge density of
mobile charge carriers. }

\subsection{Boltzmann Transport and Q1D Fermi Surface}
Simple Drude calculations are not capable of handling the highly
anisotropic nature of the Bechgaard salts nor the angular orientation
of the field relative to the lattice vectors.
The simplest treatment which includes the bandstructure comes from
a steady state
Boltzmann equation. It has previously been shown that Boltzmann transport,
appropriately modified to follow electron trajectories over many
Brillouin zones (or equivalently many Umklapp scatterings) can give
magic angle effect in \rzz\. Moreover, this model is in qualitative
agreement with the measurements in the \clo4 salt.
Although, Boltzmann transport and the Osada model have not been
successful for \pf6, they are still the only reasonable single particle
treatment available. We therefore performed both numerical and analytic
calculations using a Boltzmann Transport formulation based on the Tight
Binding Approximation band structure within the single Relaxation Time
Approximation. For simplicity,the triclinic crystal structure of
\tmt\ is taken as orthorhombic.
\be
\varepsilon=-2t_a\cos k_xa-2t_b\cos k_yb-2t_c\cos k_zc
\label{e:disp}
\ee
For \tmt, $t_a\gg t_b \gg t_c$. Often people linearize the $k_x$
dispersion for simplicity.  One would obtain the so-called linearized
dispersion:
\be
\varepsilon-\varepsilon_f=\pm\hbar v_f(k_x\mp k_f)
-2t_b\cos(k_yb)-2t_c\cos(k_zc)
\label{e:disL}
\ee
In this approximation, the Fermi velocity $v_f$ (or
the density of states on Fermi surface $N_f(\varepsilon_f)=1/\hbar v_f$)
is a constant for a given energy. Many of the angular magneto-resistance
oscillations (AMRO), \eg\ the Danner-Kang-Chaikin oscillation
(\va\vc-rotation)\cite{Danner1994}, the third angle effect
(\va\vb-rotation)\cite{Osada1996} and the combination of
them\cite{Lee1998a} can be understood within this approximation.
There is excellent agreements between experiment and theory, especially
for \clo4. However, for a linearized dispersion relation the Hall effect
is zero($\sigma_{xy}=\sigma_{xz}=0$).
It is not surprising the Nernst effect is also zero ($S_{xy}=S_{xz}$=0)
in this approximation. In order to calculate the
Nernst effect, we have to use the full dispersion or a nonlinear
approximation in either numerical computation or
in an analytic calculation.

In general transport theory, we consider both an electric current $\vec{J}$
and a thermal current $\vec{J}_q$ in response to an
electric field $\vec{E}$, a magnetic field $\vec{B}$
and a temperature gradient $(-\grad{T})$:
\begin{equation}
\left\{\begin{array}{l}
\vec{J}=\ts{\sigma}\cdot\vec{E}+\ts{\alpha}\cdot(-\grad{T}) \\

\vec{J}_q=T\ts{\alpha'}\cdot\vec{E}+\ts{\kappa}\cdot(-\grad{T})
\end{array}\right.
\label{e:J}
\end{equation}
Here $\ts{\sigma}$ is the electric conductivity tensor,
$\ts{\alpha}$ and $\ts{\alpha'}$ the thermoelectric tensor
and $\ts{\kappa}$ is the thermal conductivity tensor.
Here $\alpha_{ij}(\vec{H})=\alpha'_{ji}(-\vec{H})$ according to Onsager
relation. \cite{Onsager1931a, Onsager1931b} For free electron gas,
these coefficients can be obtained by applying the relaxation time
approximation:\cite{mermin}
\begin{equation}
\left\{\begin{array}{l}
\ds\ts{\sigma}
=\frac{e^2\tau}{4\pi^3}
\int_{_{\varepsilon=\mu}}\frac{dS_k}{\hbar v}\ \vec{v}\bk{\vec{v}}\\

\ds\ts{\alpha}=\frac{1}{eT}\frac{\pi^2}{3}(k_{_B}T)^2
\left.\pp{\ts{\sigma}(\varepsilon)}
{\varepsilon}\right|_{_{\varepsilon=\mu}}
\end{array}\right.
\label{e:sigma}
\end{equation}
Here we assume an energy and momentum independent relaxation time $\tau$.
The velocity average is defined as:  $\ds
\bk{\vec{v}(\vec{k}(t))}\equiv\int^0_{-\infty}\frac{dt}{\tau}\
e^{t/\tau}\vec{v}(\vec{k}(t))$, where $\vec{k}(t)$ is the semi-classical
motion of electrons on Fermi Surface in the presence of a magnetic field.
Eqs. \ref{e:sigma} are the starting point of our calculation.
From the general transport equations (Eqs. \ref{e:J}), by setting
$\vec{J}=0$ as the boundary condition in thermoelectric
measurements, one would find Eq.\ref{e:S}.
The Nernst signal \nzx\ is an off-diagonal element of
thermoelectric power tensor $\ts{S}$.

Now let's focus on a quasi-1D (Q1D) system, (TMTSF)$_2$X. The
Fermi surface of Q1D consists of two slightly warped sheets. In
order to evaluate the Fermi surface averaging velocity
$\bk{\vec{v}}$ we have to calculate the motion of electrons on
Fermi surface in the presence of the magnetic field.  Here the
\va-axis ($k_x$) is the best conductivity
 direction, so the Fermi surface is approximately normal to the $k_x$ axis.
Therefore,
$k_x$ is a function of $k_y$, $k_z$ and $\varepsilon$ from the dispersion
relation (Eq. \ref{e:disp}), \ie
\begin{equation}
k_x=k_x(k_y, k_z; \varepsilon)
\label{e:kx1}
\end{equation}
Therefore,
$k_x$ is not an independent variable for the semiclassical motion
of electrons, given that the motion of
the election is confined to the Fermi surface in the presence of
the magnetic field \vec{B}.
The Equations of motion can be reduced to:
\begin{equation}
\left\{
\begin{array}{l}
\ds\dd{k_y}{t}=\frac{e}{\hbar}(v_zB_x-v_xB_z) \\ \\

\ds\dd{k_z}{t}=\frac{e}{\hbar}(v_xB_y-v_yB_x)
\end{array}
\right.
\label{e:eom}
\end{equation}
By solving these two equations of motion Eq.\ref{e:eom}, we can evaluate
$\vec{v}(t)=\vec{v}(\vec{k}(t))$.  This is what we need for the
conductivity tensor $\sigma$.
>From Eqs. \ref{e:sigma}, the thermoelectric
coefficient tensor $\alpha$ is proportional to the energy
derivative of the conductivity tensor $\sigma(\varepsilon)$ at the
Fermi energy.
\be
\left\{\begin{array}{l}
\ds\sigma_{ij}=\frac{e^2\tau}{4\pi^3}
\int\!\!\!\int\frac{dk_ydk_z}{\hbar |v_x|}\ v_i\bk{v_j} \\ \\
\ds\alpha_{ij}=\frac{k_{_B}^2e\tau T}{12\pi}\ppe
\int\!\!\!\int\!\frac{dk_ydk_z}{\hbar |v_x|}\ v_i\bk{v_j}
\end{array}\right.
\label{e:sigma1}
\ee
If we could find an analytic form
of $\sigma(\varepsilon)$, evaluating $\alpha$ would be straightforward.
In our analytic calculation, we derive  an approximate analytic
form for $\sigma(\varepsilon)$, and we obtain $\alpha$ by
taking differentiation.  On the other hand, it is straightforward to
evaluate both $\sigma$ and $\alpha$ numerically. Here
$\alpha$ can be calculated by taking the energy derivative of each term
in the integral.(Eq.\ref{e:sigma1})

\subsection{Numerical Calculation}
From Eq. \ref{e:sigma1} we can obtain:
\begin{eqnarray*}
\ts{\alpha}&=&\frac{k_{_B}^2e\tau T}{12\pi\hbar}\int\!\!\!\int\!dk_ydk_z \\
&\times &\left[\underbrace{\ppe\left(\frac{1}{|v_x|}\right)\
\vec{v}\bk{\vec{v}}}_{(a)}+\underbrace{\frac{1}{|v_x|}\
\pp{\vec{v}}{\varepsilon}\bk{\vec{v}}}_{(b)}+\underbrace{
\frac{1}{|v_x|}\ \vec{v}\ \pp{\bk{\vec{v}}}{\varepsilon}}_{(c)}
\right]
\label{e:alp}
\end{eqnarray*}

Note here $\alpha\propto T$ if we assume $\tau$ is T independence,
which is a good approximation at low temperature. It is
straightforward to evaluate the first two terms.   Since:
\be
\ds \ppep{\vec{v}}=\hbar\cdot
\ts{M}^{-1}\cdot\ppep{\vec{k}}
\label{e:pvpe}
\ee
Here we define the inverse mass tensor $\ts{M}^{-1}$ as:
\be
\invm_{ij}=\frac{1}{\hbar}\pp{v_i}{k_j}
=\frac{1}{\hbar^2}\frac{\partial^2\varepsilon(\vec{k})}
{\partial k_i \partial k_j}\label{e:invm}
\ee

Here $k_y$ and $k_z$ are independent variables, so the energy
dependence of the first two terms only comes from $k_x$,
which is a function of $\varepsilon$ (see Eq.\ref{e:kx1}).
Then, we will have:
\begin{equation}
\ppe=\pp{k_x}{\varepsilon}\ppkx=\frac{1}{\hbar v_x}\ppkx
\end{equation}
Therefore:
$\ds\ppe\left(\frac{1}{v_x}\right)= -\frac{m_{_{xx}}^{-1}}{v_x^3}$
and $\ds\ppep{v_i}=\delta_{i,x}\frac{m_{_{xx}}^{-1}}{v_x}$.
For the (c) term, we exchange
the differentiation and averaging(integral), \ie
$\ds  \ppep{\bk{\vec{v}}}=\bk{\ppep{\vec{v}}}$.
Putting the (a), (b) and (c) terms together, we obtain an expression
for the thermoelectric coefficient tensor $\alpha$:
\be
\alpha=\alpha^{(+)}+\alpha^{(-)}
\ee
where $\alpha^{(\pm)}$ is defined as:
\begin{eqnarray*}
\alpha_{_{ij}}^{(\pm)}&\equiv&\frac{k_{_B}^2e\tau T}{12\pi\hbar}
\ds\int\!\!\!\!\int\! dk_ydk_z
(\pm)\left[-\frac{m_{_{xx}}^{-1}}{v_x^3}v_{_i}\bk{v_{_j}}
\right. \\ && \left.
+\frac{m_{_{xx}}^{-1}}{v_x^2}\delta_{i,x}\bk{v_{_j}}
+\frac{v_{_i}}{v_x}\bk{\ppep{v_{_j}}}\right]
\qquad k_x\stackrel{>}{<}0
\label{e:alpha_n}
\end{eqnarray*}
Now we need to find $\ds \pp{\vec{v}}{\varepsilon}(t)$,
which involves the motion of an electron in the magnetic field.
$\ds\ppep{\vec{v}}(t)$ depends on the value of the Fermi Energy not only
through $k_x$, but also through $k_y$ and $k_z$,  because
$k_y$ and $k_z$ are also functions of $t$ when the electron is moving
on the Fermi surface in the magnetic field. These functions
depend on energy $\varepsilon$ and the ``initial'' condition
($k_y^0$, $k_z^0$), \ie
\begin{equation}
\left\{\begin{array}{l}
k_y(t)=k_y(t;\varepsilon, k_y^0, k_z^0)\\ \\
k_z(t)=k_z(t;\varepsilon, k_y^0, k_z^0)
\end{array}
\right.
\end{equation}
\figepss{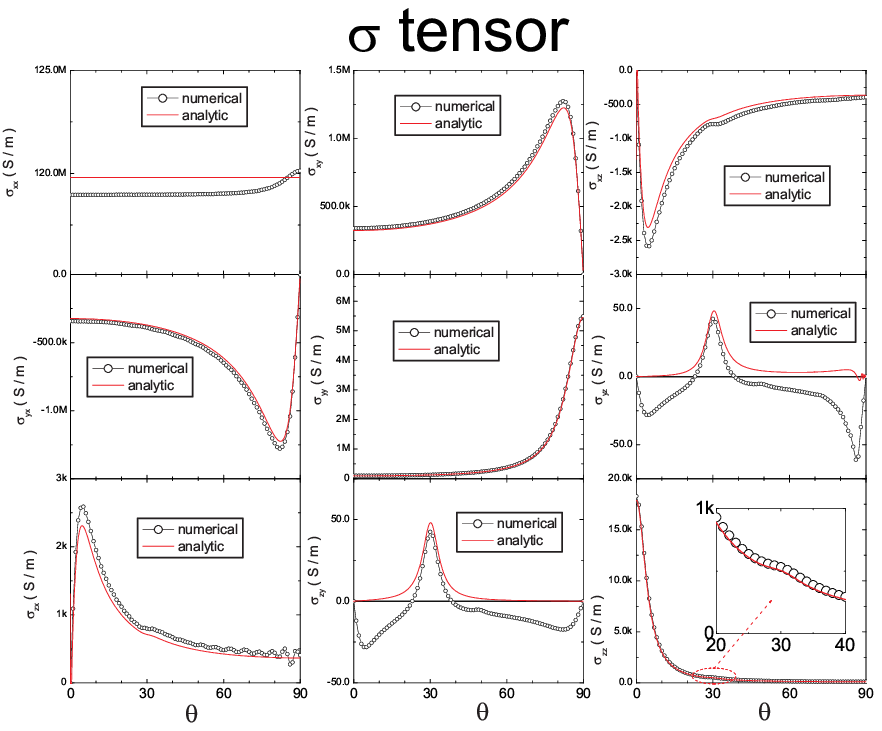}{6}{Angular dependence of conductivity tensor $\sigma$.
Here $\vec{B}$=8 Tesla. The origin is the \vc-axis, while 90\dg is \vb-axis
and 30\dg is the +1 Lebed angle: \vc+\vb. The inset of $\sigma_{zz}$ shows
a hump at the +1 Lebed angle. The open circles are the numerical calculations,
and the thin lines are the analytic results.}

We can't derive the energy dependence of $k_y$ and $k_z$ directly
if we don't know the solution of the Equations of motion.
However, we can find the differential equations which are satisfied
by $\ds\pp{k_y}{\varepsilon}$ and $\ds\pp{k_y}{\varepsilon}$,
from the equations of motion of $k_y$ and $k_z$ (Eq.\ref{e:eom}).
\begin{equation}
\left\{\begin{array}{l}
\ds\ddt\ppep{k_y}=\ppe\dd{k_y}{t}=
\frac{e}{\hbar}\left(\ppep{v_z}B_x-\ppep{v_x}B_z\right) \\ \\
\ds\ddt\ppep{k_y}=\ppe\dd{k_z}{t}=
\frac{e}{\hbar}\left(\ppep{v_x}B_y-\ppep{v_y}B_x\right)
\end{array}
\right.
\label{e:pkpe}
\end{equation}
From (Eq.\ref{e:kx1}), we will find:
\begin{eqnarray*}
\ppep{k_x(t)}
&=&\frac{1}{\hbar v_x(t)}\left[1-\hbar v_y\ppep{k_y(t)}
-\hbar v_z\ppep{k_z(t)}\right]
\end{eqnarray*}
Combining Eq.\ref{e:pkpe} with Eq.\ref{e:pvpe}, we can
obtain numerical solutions of $\ds\ppep{\vec{v}}(t)$.
Since $k_y$, $k_z$ are independent of
$\varepsilon$ at $t=0$,
the ``initial'' conditions for $\ds\ppep{k_y}(t)$ and
$\ds\ppep{k_z}(t)$ are $\ds\pp{k_y}{\varepsilon}(0)
=\pp{k_z}{\varepsilon}(0)=0$.

\figepss{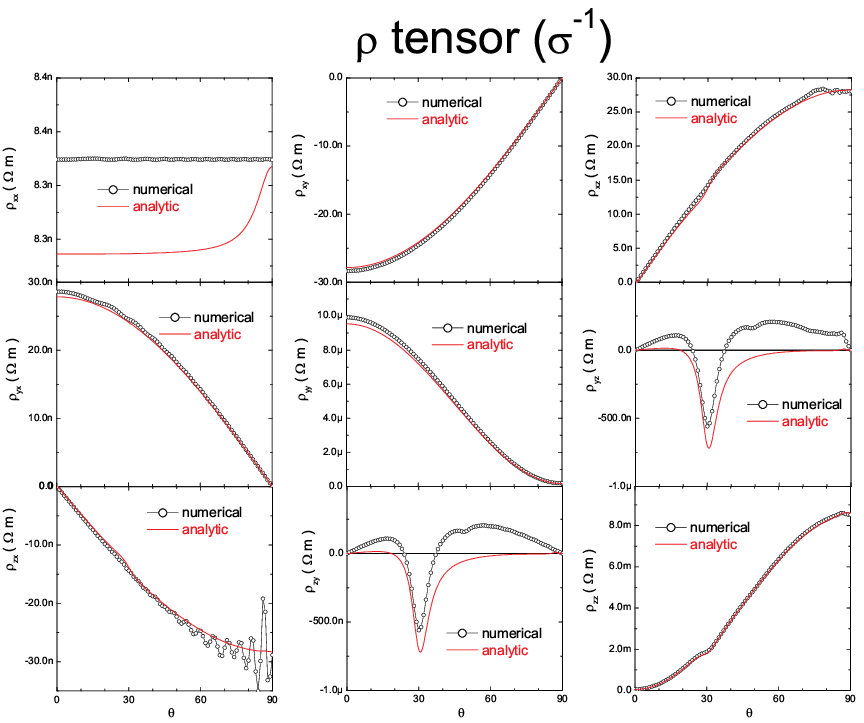}{6}{Resistivity tensor $\rho=\sigma^{-1}$.
Here we just perform a simple matrix inverse operation.}

In summary, to calculate both $\ts{\sigma}$ and $\ts{\alpha}$,
we numerically
solve two sets of equations of motions, Eqs. \ref{e:eom} and
Eqs. \ref{e:pkpe}.
To treat the differential equations in Eqs. (\ref{e:eom} and
\ref{e:pkpe}), we use a 4th order Runge-Kutta Method.\cite{Recipes}
Then we numerically integrate $\bk{\vec{v}}$ and $\ds\bk{\pp{\vec{v}}
{\varepsilon}}$ and evaluate both $\sigma$ and $\alpha$
by Fermi surface integrals.  Here we use the band parameters
($t_a=0.25 eV$, $t_b=0.024 eV$ and $t_c=0.008 eV$)
from tight binding approximations,
realistic lattice parameters (a=3.49 $\AA$, b=7.7 $\AA$ and
c=13.264 $\AA$) and a scattering time
$\tau=4.26\times 10^{-12}$\emph{sec} from previous studies
by Danner \etal\cite{Danner1994}.  Here we use B=8 Tesla,
and T=1 K, which are comparable with experiment conditions.
The combination of the parameters produce
$\ds\omega_c\tau=\frac{eB\tau}{m_e}=6$ for an
isotropic free electron gas.
We use a $20\times 20$ grid on the Fermi surface in the calculations.
Although it is a little rough, it catches all the main features.
We also used a $40\times 40$ grid for some points and did not find a
significant difference.

\subsection{Analytic Calculation}
The basis of our analytic approximation scheme is
finding the proper correction to the linear dispersion
approximation Eq.\ref{e:disL}.
We expanded $v_x$ to next
order to include the effect of non-linearity of the dispersion,
\ie\ $v_x\approx v_f+\delta v_f$. This
approach is basically the semi-classical version of Lebed's quantum
approach\cite{Lebed2004}. It is straightforward to find:
$\ds v_f=\frac{2t_aa}{\hbar}\sin k_fa\quad \mbox{here}\quad\cos
k_fa=-\frac{\varepsilon}{2t_a}\quad \mbox{and}\\
\ds \frac{\delta v_f}{v_f}=\frac{\cos
k_fa}{\sin^2k_fa}\left[\frac{t_b}{t_a}\cos(k_yb)
+\frac{t_c}{t_a}\cos(k_zc)\right]+...$

Defining $\ds\beta\equiv-\frac{\cos k_fa}{\sin^2k_fa}\frac{t_b}{t_a}$
and $\ds\gamma\equiv-\frac{\cos k_fa}{\sin^2k_fa}\frac{t_c}{t_a}$,
then we have:
\be
\left\{\begin{array}{l}
\ds v_x\approx v_f[1-\beta\cos(k_yb)-\gamma\cos(k_zc)]\\
\\
\ds v_y=\frac{2t_bb}{\hbar}\sin k_yb \\
\\
\ds v_y=\frac{2t_cc}{\hbar}\sin k_zc
\end{array}
\right.
\label{e:vyvz}
\ee
By substituting $v_x$ in the equations of motion Eqs.\ref{e:eom},
one can obtain analytic expressions of $k_y(t)$ and $k_z(t)$
and evaluate velocity averages $\bk{\bf v}$ for $\sigma(\varepsilon)$.
Then it is straightforward to obtain $\alpha$ from Eq.\ref{e:sigma}.
Details of the analytic calculation are presented in the appendix.
\figepss{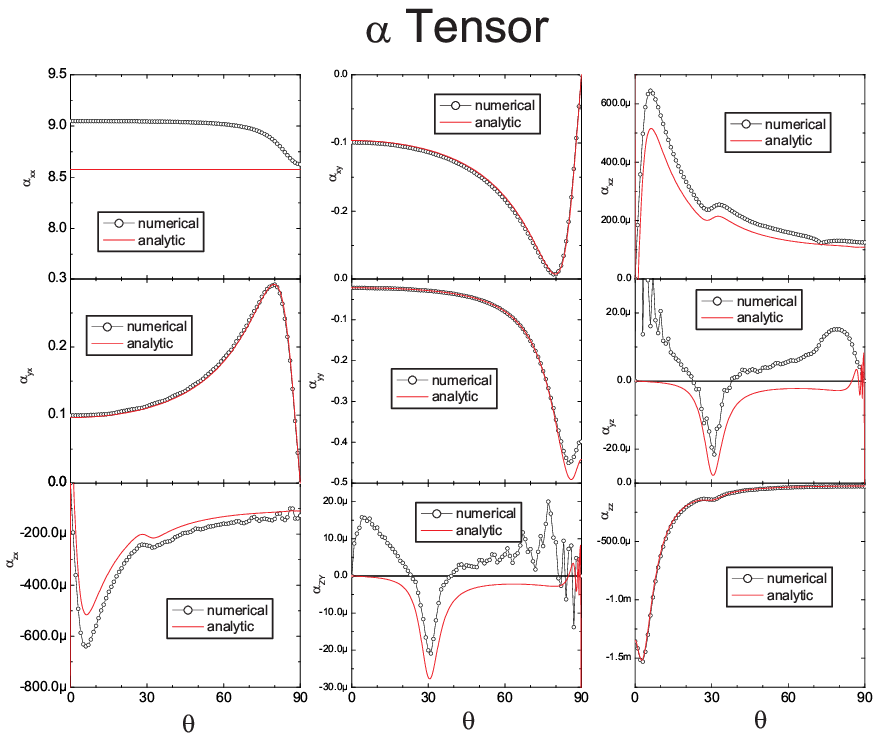}{6}{Thermoelectric coefficient tensor $\alpha$.
In Boltzmann transport, $\ds\alpha\sim\left.\pp{\sigma}{\varepsilon}
\right|_{\mu}$.}
As shown in Fig.\ref{sigmaT}, Fig.\ref{rhoT}, Fig.\ref{alphaT} and
Fig.\ref{thpT}, our analytic calculation reproduces the main features
of the numerical calculation.
In some cases, the results from different methods overlap.
Therefore, we believe our calculations describe the main behavior of the Nernst
signal in Boltzmann transport within the tight binding approximation.

\subsection{Results}
Fig.\ref{sigmaT} shows the calculated angular dependence of the
conductivity tensor $\sigma$.  The graphs are arranged in the
pattern of the tensor elements $\sigma_{ij}$ in the matrix form.
Here $\theta$ is defined respect to the \vc-axis. Therefore $\theta=$90\dg
corresponds to the magnetic aligning at the \vb-axis.
As shown in the insert graph of $\sigma_{zz}$, there is a small hump
at 1st Lebed angle: \vc+\vb, which is about 30\dg in our approximation.
This is one test that our calculations reproduce the angular dependence of
$\sigma_{zz}$ calculated by Osada \etal\cite{Osada1999}. By increasing the
scattering time $\tau$, or magnetic field, we can clearly resolve a peak
at this angle. We also confirmed other AMROs, \eg\ \va\vc-rotation and
reproduce the Danner oscillations\cite{Danner1994}.

However, as far as we know, there is no calculation of tensor $\alpha$ in
the literature for comparison.  This was our motivation for performing
the analytic calculation to confirm our numerical results. Once we
obtained $\sigma$ and $\alpha$,
we got the thermoelectric power tensor $\ts{S}$ (Fig.\ref{thpT})
by taking the product of $\rho=\sigma^{-1}$ (Fig. \ref{rhoT})
and $\alpha$ (Fig. \ref{alphaT}).

It is clear that the quality of numerical calculation of $\sigma$
is much better than that of $\alpha$.
For $\sigma$ most curves are very smooth and only minor oscillations are
observed. For $\alpha$ most curves are smooth, except $\alpha_{yz}$ and
$\alpha_{zy}$.  $\alpha_{yz}$ has some spiky features for $\theta$ close to
\vc; while $\alpha_{zy}$ has some spiky features for $\theta$ close to \vb.
This is because the energy derivative of the velocities is very sensitive
to the location on the Fermi surface ($k_y$, $k_z$).  Finite size grid
integration
could also generate artificial spikes if the integrand oscillates a lot.

By varying the grid size and the integration cut off limit, these artificial
features can be suppressed, but at the expense of much more computation
time.  Since we are only interested in the general behaviors and magnitudes
for a given set of parameters, we will use these non-perfect calculation
results to compare with experiments,  while keeping in mind that sharp features
might be artificial. Also, our analytic results will help us to find out the
physical features.

\figepss{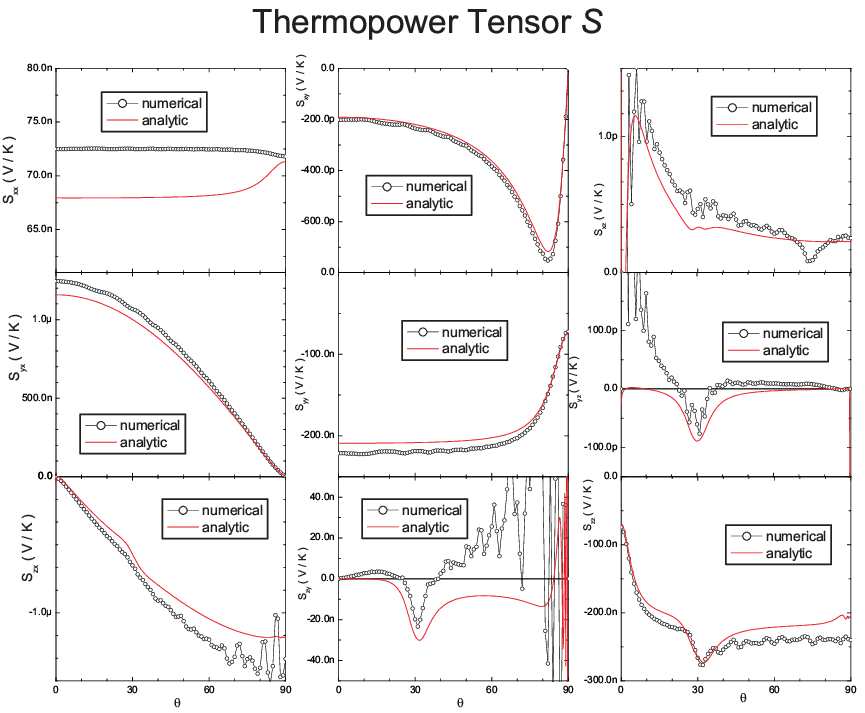}{6}{Thermoelectric Power tensor $\ts{S}$,
which is the product of resistivity tensor $\rho$(Fig.\ref{rhoT})
and thermoelectric coefficient tensor $\alpha$ (Fig.\ref{alphaT}).}

The Nernst signal $S_{zx}$ corresponds to the experimental results discussed
in the previous section. It is clear that its angular dependence is
$\sin\theta$-like, which agrees with the simple Drude model. The maximum
magnitude is about 1 $\mu$V/K, which is 2 orders smaller than
what we found in \pf6 as shown in Fig. \ref{calexp}.
This result is very different in shape from our observation, missing
the resonances at magic angles and it gives the wrong
temperature dependence ($\ts{S}\propto T$ in Boltzmann Transport).
Therefore, we conclude that the single particle picture is not able to
describe the Nernst effect observed in \pf6.

It is interesting to note that the Nernst signal $S_{xz}$ has a
similar angular dependence as the experimental Nernst effect
$S_{zx}$, showing a peak near \vc\ and a sign change. However, the
geometry is completely opposite and the value is about 8 orders of
magnitude too small. Experimentally we couldn't detect a sizable
$S_{xz}$,  though we did not optimize the experimental setup for
that measurement. \figeps{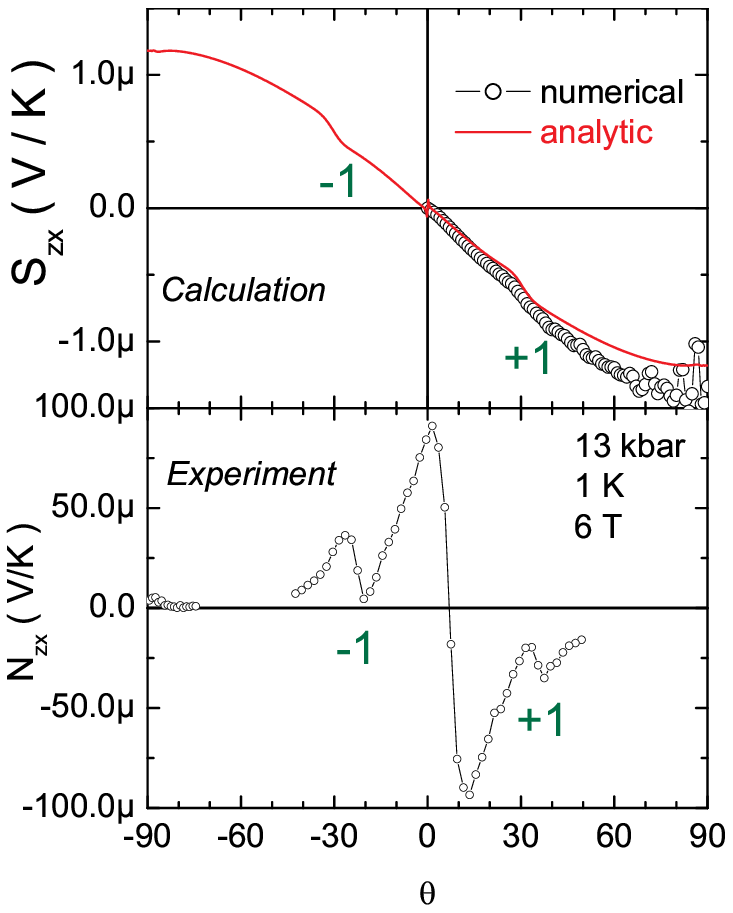}{3}{Comparison of \nzx between
calculations (Fig. \ref{thpT}) and experimental data (Fig.
\ref{NRQPcom}).} Comparing the off-diagonal elements of $\ts{S}$
(\ie\ the Nernst effect), we find that the elements in the upper
triangle, $S_{xy}$, $S_{xz}$ and $S_{yz}$ are much smaller that
those in the lower triangle, $S_{yx}$, $S_{zx}$ and $S_{zy}$.  More
surprisingly, the conjugate elements don't have the same angular
dependence. Does this violate the Onsager
relation\cite{Onsager1931a,Onsager1931b}? The answer is no. The
Onsager relation only states constraint on $\sigma$ and $\alpha$ and
$\alpha'$ (defined in Eqs. \ref{e:J}): \be \left\{\begin{array}{l}
\sigma_{ij}(H)=\sigma_{ji}(-H)\\
\alpha_{ij}(H)=\alpha'_{ji}(-H)
\end{array}\right.
\label{e:onsager}
\ee

By symmetry, we know that $\sigma_{ij}$ ($\alpha_{ij}$) is an odd
function of magnetic field $\vec{H}$ for $i\neq j$. \be
\left\{\begin{array}{l}
\sigma_{ij}(H)=-\sigma_{ij}(-H)\\
\alpha_{ij}(H)=-\alpha_{ij}(-H)
\end{array}\right.
\label{e:odd}
\ee

Together with the Onsager relation Eq.\ref{e:onsager}, we can
find: \be \left\{\begin{array}{l}
\sigma_{ij}(H)=-\sigma_{ji}(H)\\
\alpha_{ij}(H)=-\alpha_{ji}(H)
\end{array}\right.
\label{e:off}
\ee

This is exactly what we see in the calculations (except $\sigma_{yz}$
($\alpha_{yz}$) and $\sigma_{zy}$($\alpha_{zy}$), which are not real
Hall effects since the magnetic field is in the plane).
However, the thermopower
tensor $\ts{S}$ is the product of $\rho=\sigma^{-1}$ and $\alpha$. In
general one should not expect $S_{ij}=-S_{ji}$.  This is only true
when we consider an isotropic system, where
$\rho_{ii}=\rho_{\circ}$ and $\rho_{ij}=\rho_H$ for $i\neq j$. In
an anisotropic system like \tmt, $\rho_{xx}\ll\rho_{yy}\ll\rho_{zz}$.
If we ignore the Hall effect, we will have
$S_{xy}\sim\rho_{xx}\alpha_{xy}\ll\rho_{yy}\alpha_{yx}\sim
S_{yx}$.  Nevertheless, the Nernst effect $S_{ij}$ ($i\neq j$)is an odd function of
$\vec{H}$ and Seebeck effect $S_{ii}$ is an even function of $\vec{H}$ as long as
there is a inversion symmetry.

In summary, we numerically and analytically calculated the thermopower tensor
$\ts{S}$ by evaluating both the conductivity tensor $\sigma$ and
thermoelectric coefficient tensor $\alpha$.  The numerical results
agree well with the analytic approximation.  This gives us confidence
on the reliability of our calculations.
It is clear that Boltzmann transport
within a single particle picture is not consistent with our observation in
\pf6.
Therefore, correlation effects due to the strong \emph{e-e} interaction
should be considered in understanding the giant Nernst effect
found in \pf6.
However, we note that the angular dependence of $S_{zx}$ fits the
data in \clo4 very well, though there is a factor of 10 or so difference
in magnitude.\cite{Nam2005}
Our results are not limited to the Bechgaard salts \tmt.  For any Q1D
system with open Fermi surface, all the transport coefficients can be
calculated using our results based on Boltzmann transport in a tight
binding model. Our original results should prove useful for further
investigations.

\section{Discussion}
The giant value of the Nernst effect and the Nernst resonances at Magic angles are
 not understood and it appears
difficult to explain them in conventional Fermi Liquid models as illustrated by
comparing our experiments and Boltzmann transport calculations.
An exotic feature is that the Nernst signal changes its sign sharply
at magic angles, with 3 ``resonances'' within 70$^{\circ}$ in our
measurements.  As we know, the sign of the transverse electric field is determined by
the cross product: $\nabla T \times {\bf B}$. Of course, the physics
really involves $\vec{E}=\vec{v}\times\vec{B}$. Since the temperature
gradient is fixed, the only quantity that could possibly change
its sign is the component of {\bf B}.
As the magnetic field passes through a magic angle,
the only component of magnetic field that could change sign
relative to a magic angle is the one that is perpendicular to the direction
of the magic angle.  Therefore, we have to conclude that the Nernst signal
in \pf6 comes from $\vec{v} \times {\rm B}_{\perp}$. This means that
 whatever is moving is confined in the Magic angle plane. The main idea that
underlies our interpretation of the Nernst resonances is that the
transport is only coherent at the planes which are ``parallel''(or
close to parallel) to the magnetic field.  In other words, the
coherent electronic motion is controlled by the orientation of the
magnetic field relative to the planes defined by the conducting
chains and the interchain directions. (Fig.\ref{coh_plane})

\figeps{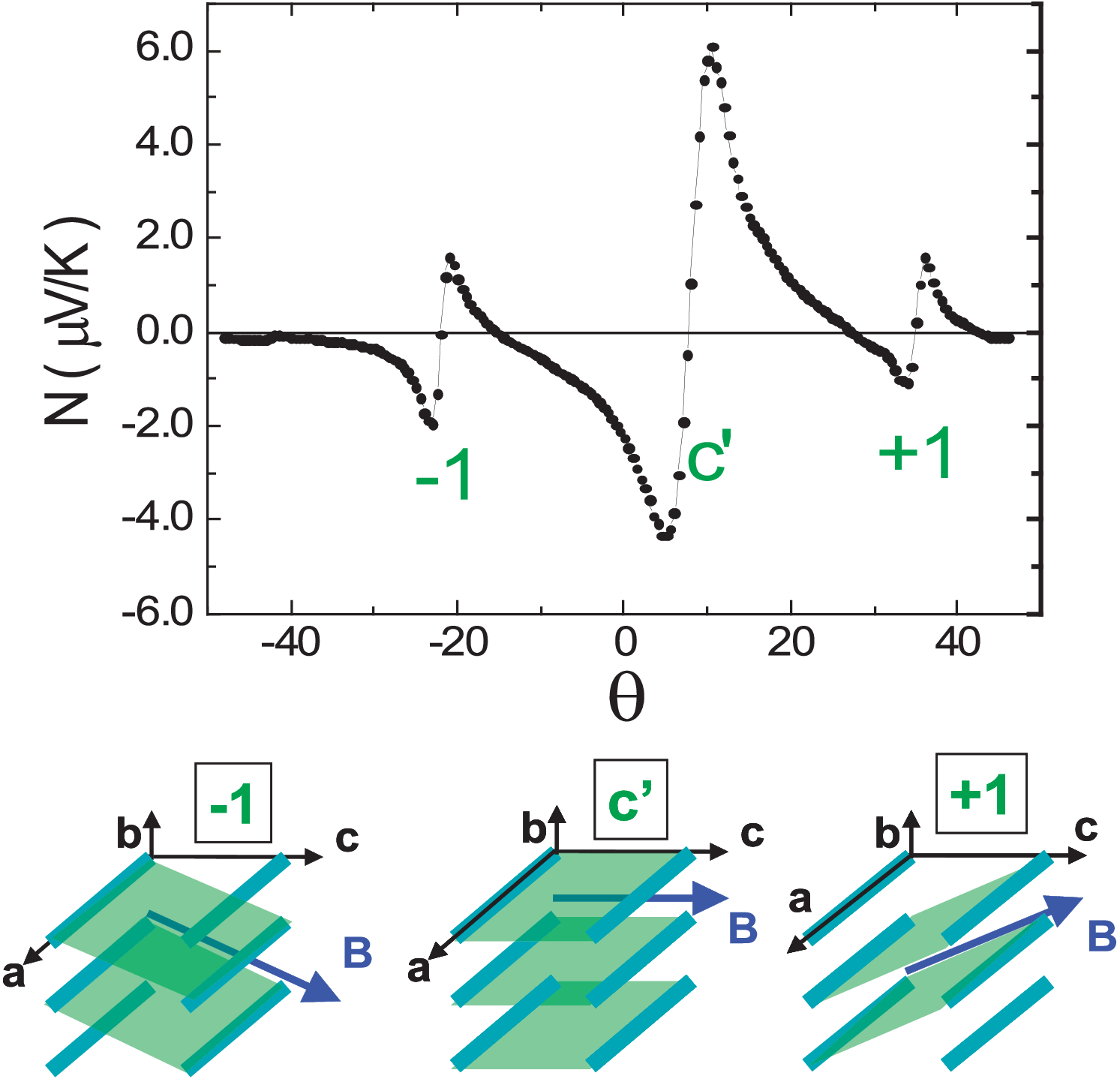}{3}{The schemaic picture indicates Coherent
Transport only in the magic angle planes which are parallel to
magnetic field. The Nernst signal would chagne sign as the field is
rotated above or below these planes. \fbox{-1}: ${\bf B}\parallel
\mbox{-1L}$, \va(\vc'-\vb')-plane; \fbox{\vc'}: ${\bf B}\parallel
\mbox{\vc'}$, \va\vc'-plane; \fbox{+1}: ${\bf B}\parallel
\mbox{+1L}$, \va(\vc'+\vb')-plane. The Nernst data is taken from
previous measurements\cite{Wu2003}.}

The nature of the coherence is not clear at this moment.
One possibility is quasi-particle coherence.
For example we may have a \emph{Field Induced Interchain/Interplane
Decoupling} picture.

\figeps{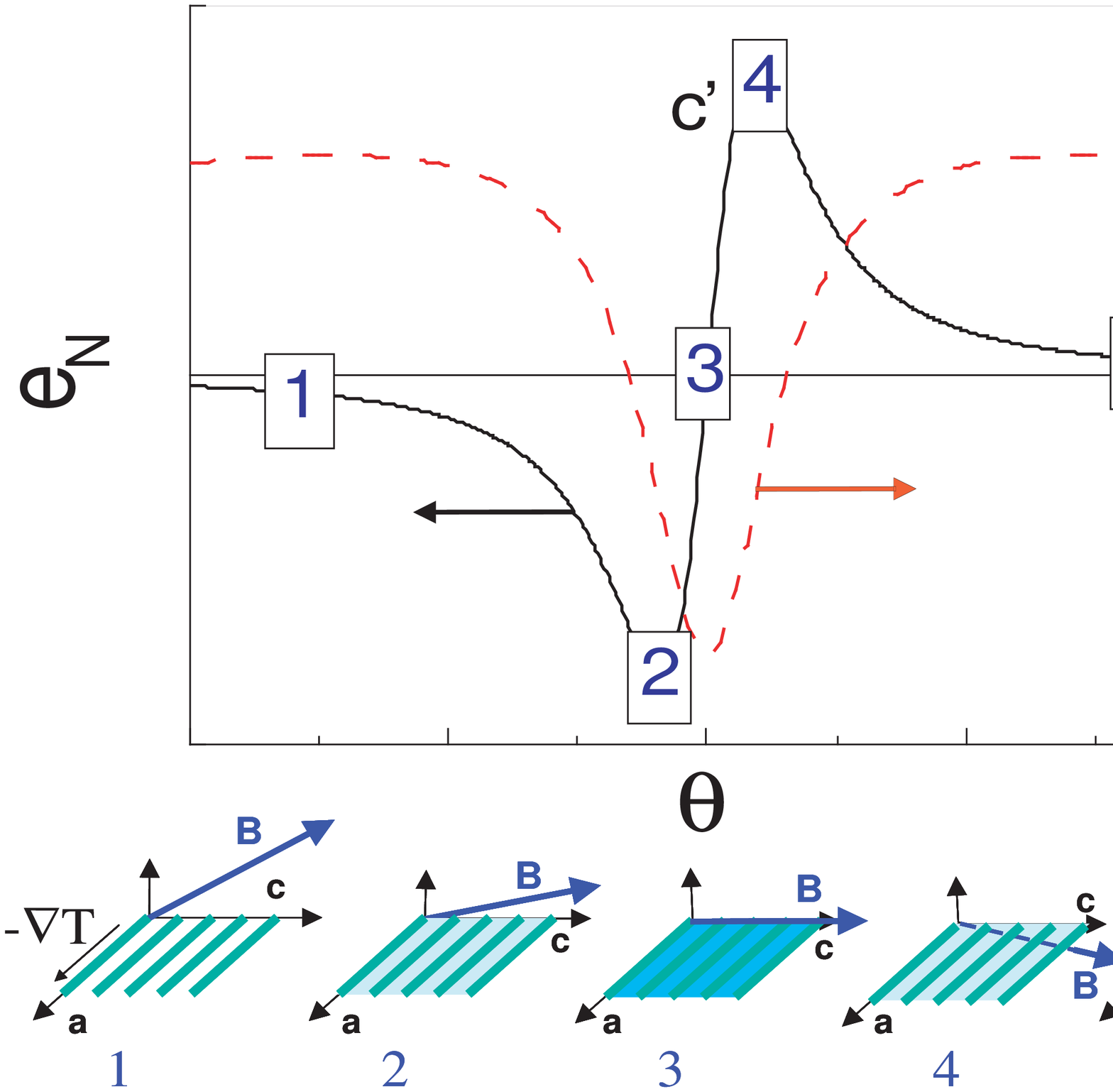}{3}{\emph{Field Induced Interchain/Interplane Decoupling}\\
\fbox{1}: ${\bf B}_{\perp}>{\bf B}^*$,
$t_c^{\mbox{eff}}=0$,
chains are decoupled, Nernst signal is zero $e_N=0$;\\
\fbox{2}: ${\bf B}_{\perp}<{\bf B}^*$, $t_c^{\mbox{eff}}\neq
0$,
chains are coupled,  $e_N>0$;\\
\fbox{3}: ${\bf B}_{\perp}=0$, $t_c^{\mbox{eff}}\neq 0$,
chains are coupled,  $e_N=0$;\\
\fbox{4}: ${\bf B}_{\perp}<{\bf B}^*$, $t_c^{\mbox{eff}}\neq
0$,
chains are coupled,  $e_N<0$;\\
\fbox{5}: ${\bf B}_{\perp}>{\bf B}^*$, $t_c^{\mbox{eff}}=0$,
chains are decoupled, Nernst signal is zero $e_N=0$.}

Fig.\ref{FIICD} shows the basic idea of this picture. \pf6 is a Q1D
system, which consists of conduction chains along the \va-axis.  The
chains are weakly coupled with each other. When the magnetic field
{\bf B} is far from a magic angle, \eg\ \vc', the inter-chain
coherent coupling in this direction is effectively suppressed by the
large normal field component, \ie, $t_c^{\mbox{eff}}=0$ when ${\bf
B}_{\perp}>{\bf B}^*$ (here \textbf{B}$^*$ is a cross-over magnetic
field scale). In other words, there is no coherent transport in the
\va\vc-plane for sufficient perpendicular field. (Fig.\ref{FIICD}
\fbox{1}, \fbox{5}\ ) When {\bf B} is parallel to the \vc'
direction, the coherent coupling in the \vc-direction is restored
and a coherent \va\vc-plane is formed. (Fig.\ref{FIICD} \fbox{3}\ )
If we tilt the magnetic field slightly upward, there is a small
component ${\bf B}_{\perp}$ of {\bf B} normal to this
plane.(Fig.\ref{FIICD} \fbox{2}\ ) Here $\delta\theta$ is small
enough that ${\bf B}_{\perp}<{\bf B}^*$. So the plane is still
coherent. In the presence of temperature gradient $-\nabla T$, there
is a transverse electric field {\bf E}, the Nernst effect. When we
tilt the field slightly to the other side of the magic angle,
everything is the same except the sign of ${\bf B}_{\perp}$
reverses.  Therefore, the sign of the Nernst signal is reversed.
(Fig.\ref{FIICD} \fbox{4}\ ) This scenario is then repeated at the
other magic angles as in Fig.\ref{coh_plane}.

In fact, the idea of field induced decoupling is not new.
Strong \etal\cite{Strong1994} considered the isolated conducting planes,
\va\vb-planes, of \pf6 as 2D Non-Fermi-Liquid due to the strong \emph{e-e}
interactions. The possible non Fermi Liquid ground state of "isolated"
\pf6 chains is supported by transport\cite{Danner1995}, optical
\cite{Vescoli1998} and thermal transport\cite{Lorenz2002}
measurements. The effect of field induced
coherent coupling/decoupling is supported by temperature
and angular dependence magneto-resistance
studies\cite{Katya_thesis1998}. However,
this theory is not universally accepted due to the unknown nature of the
non Fermi Liquid state and the lack of a detailed model.

Another possibility is the
2 dimensional superconducting phase coherence proposed by
Ong \etal\cite{Ong2004}.  Ong points out that normal
quasi-particles usually give a thermopower much larger than the Nernst
signal,\cite{Wang2001} whereas we have a large Nernst signal with undetectably
small thermopower.
On the contrary, vortex flow naturally produces an electric
field that is predominantly transverse.
This is generally true for most conventional systems,
as well as HT$_{c}$ cuprates.\cite{Wang2001}
Implicit in this model is the ability of the magnetic
field to destroy phase coherence in the planes to which it is normal.
It is natural that the vortices penetrating perpendicular planes destroy
superconductivity. This is similar to the decoupling model discussed above
 and has the consequent"resonances" at magic angles.

\section{Conclusion}
We present a detailed study of the Nernst effect \nzx\ in \pf6
as a function of temperature,pressure and magnetic field magnitude
and direction. The data agree well with our previous
measurements.\cite{Wu2003}  We have calculated Boltzmann
transport coefficients by both numerical and analytic methods
with realistic band parameters within the single relaxation time
approximation. The two calculations agree with either very well,
but fail to describe the experimental data. The large magnitude,
resonant-like angular field dependence
and the non-linear field and temperature dependence cannot be understood
within the semi-classical approximation of Boltzmann transport.
The sign change of the Nernst effect at magic angles
strongly suggests that the transport is effectively 2-dimensional
in lattice planes parallel or close to the orientation of magnetic field.
The nature of the coherence is not clear at this moment.
We suggest that the answer may lie in field induced decoupling
of the strongly correlated chains.

\section{Acknowledgements}
We are grateful to Phil Anderson, David Huse, Iddo Usshinski and Vadim
Ogasyan for helpful discussions.

\section{Appendix}
Define $\phi_b=k_yb$ and $\phi_c=k_zc$. In the presence of a magnetic field
in the \vb\vc-plane $\vec{B}=(0, B\sin\theta, B\cos\theta)$, define
$\omega_b^{\theta}\equiv\omega_b\cos\theta$ and
$\omega_c^{\theta}\equiv\omega_c\sin\theta$, here
$\hbar\omega_b=|e|v_fBb$ and $\hbar\omega_c=|e|v_fBc$. $e<0$ is
the electron charge.  Then the equation of motion become: \be
\left\{\begin{array}{l} \ds\dd{\phi_b}{t}=
\omega_b^{\theta}(1-\beta\cos\phi_b-\gamma\cos\phi_c)\\
\ds\dd{\phi_c}{t}=
-\omega_c^{\theta}(1-\beta\cos\phi_b-\gamma\cos\phi_c)
\end{array}
\right.
\label{e:eomc}
\ee
The exact solutions are hard to obtain.  However, since
$\beta,\gamma\ll 1$, the solutions can be approximated
to 1st order by an iterative method:
\be
\left\{\begin{array}{l}
\ds\phi_b(t)\doteq\phi^0_b+\obt t-\beta\sin(\phi^0_b+\obt t)
-\gamma\frac{\obt}{\oct}\sin(\phi^0_c-\oct t)\\
\ds\phi_c(t)\doteq\phi^0_c-\oct t
+\beta\frac{\oct}{\obt}\sin(\phi^0_b+\obt t)
-\gamma\sin(\phi^0_c-\oct t)
\end{array}\right.
\ee
Note here $\ds\frac{\obt}{\oct}$ diverges as $\theta\ra 0$,
and $\ds\frac{\oct}{\obt}$ diverges as $\theta\ra 90^{\circ}$.
So this solution is only good at $0<\theta<90^{\circ}$. One has to
be aware that as \vec{B} approaches \vb\ or \vc, this solution may
not give the correct result.  With the help of Jacobi's expansions,
it is straightforward to evaluate $\vec{v}(t)$ (Eq.\ref{e:vyvz})
and $\bk{\vec{v}}$.  Then we can obtain analytic expression for
$\sigma$.
After tedious but straightforward calculation,
we obtain the conductivity tensor $\sigma$.  Here we assume
$\gamma\ll\beta\ll 1$, and only keep the leading terms
in $\beta$, $\gamma$, $\beta^2$ and $\beta\gamma$.
For simplicity,  we also use the anti-symmetric
property of $\sigma$.  It is clear from Fig.\ref{sigmaT} that the
analytic calculation reproduces the numerical results very well.
Due to the limitation of our expansion, not every minor detail was
reproduced. For example, in this 1st order approximation,
$\sigma_{xx}$ is independence of angle in analytic form, while
numerically it shows a very weak angle dependence.
The good agreement between different calculations give us confidence
about the reliability of the calculations.
Once we know the analytic form of the conductivity tensor $\sigma$,
we can take its energy derivative to obtain $\alpha$.
\begin{widetext}
\be
\left\{\begin{array}{l}
\ds\sigma_{xx}=\frac{2e^2\tau}{\pi\hbar}
\frac{v_f}{bc}(1-\beta J_1(\beta)-\gamma J_1(\gamma))\\
\\
\ds\sigma_{yx}=-\frac{2e^2\tau t_b}{\pi\hbar^2c}
\left\{\beta\frac{\obtt}{1+(\obtt)^2}
-\frac{\beta^2}{2}J_1(\beta)\frac{2\obtt}{1+(2\obtt)^2}
+\left[\frac{\beta\gamma}{2}J_1\left(\gamma\frac{\obt}{\oct}\right)
+\frac{\gamma^2}{2}J_1\left(\beta\frac{\oct}{\obt}\right)\right]
\right.\\ \qquad \ds
\left.\times\left[\frac{(\obt-\oct)\tau}{1+(\obt-\oct)^2\tau^2}
-\frac{(\obt+\oct)\tau}{1+(\obt+\oct)^2\tau^2}\right]
\right\}
\\ \\
\sigma_{xy}=-\sigma_{yx}\\
\\ \\
\ds\sigma_{yy}=\frac{4e^2\tau t_b^2}{\pi\hbar^3v_f}\frac{b}{c}
\left\{\frac{1}{1+(\obtt)^2}+\frac{\beta J_1(\beta)}{2}
\frac{1}{1+(2\obtt)^2}\right. \\ \qquad \ds\left.
+\frac{\gamma}{2}J_1\left(\gamma\frac{\obt}{\oct}\right)
\left[\frac{1}{1+(\obtt-\octt)^2}-\frac{1}{1+(\obtt+\octt)^2}\right]
\right\}\\ \\

\ds\sigma_{zx}=\frac{2e^2\tau t_c}{\pi b\hbar^2}
\left\{\gamma\frac{\octt}{1+(\octt)^2}
+\frac{\gamma^2}{2}J_1(\gamma)\frac{2\octt}{1+(2\octt)^2}
-\left[\frac{\beta^2}{2}J_1\left(\gamma\frac{\obt}{\oct}\right)
+\frac{\beta\gamma}{2}J_1\left(\beta\frac{\oct}{\obt}\right)\right]
\right.\\ \qquad\ds
\left.\times\left[\frac{(\obt-\oct)\tau}{1+(\obt-\oct)^2\tau^2}
+\frac{(\obt+\oct)\tau}{1+(\obt+\oct)^2\tau^2}\right]
\right\}\\\\

\sigma_{xz}=-\sigma_{zx}\\ \\

\ds\sigma_{yz}=\frac{2e^2\tau t_bt_c}
{\pi\hbar^3v_f}\gamma J_1\left(\beta\frac{\oct}{\obt}\right)
\left[\frac{1}{1+(\obtt+\octt)^2}+\frac{1}{1+(\obtt-\octt)^2}\right]
\\ \\
\ds\sigma_{zy}=\frac{2e^2\tau t_bt_c}{\pi\hbar^3v_f}
\beta J_1\left(\gamma\frac{\obt}{\oct}\right)\left[
\frac{1}{1+(\obtt-\octt)^2}+\frac{1}{1+(\obtt+\octt)^2}\right]\\ \\

\ds\sigma_{zz}=\frac{4e^2\tau t_c^2}{\pi\hbar^3v_f}\frac{c}{b}
\left\{\frac{1}{1+(\octt)^2}-\frac{\gamma J_1(\gamma)}{2}
\frac{1}{1+(2\octt)^2}\right. \\ \qquad\ds\left.
+\frac{\beta}{2}J_1(\beta\frac{\oct}{\obt})
\left[\frac{1}{1+(\obtt-\octt)^2}-\frac{1}{1+(\obtt+\octt)^2}\right]
\right\}
\end{array}\right.
\ee
\be
\left\{\begin{array}{l}
\ds\alpha_{xx}=-\frac{2\pi k_B^2Te\tau a}{3\hbar^2 bc}
\{(1-\beta J_1(\beta)-\gamma J_1(\gamma))
+3[\beta J_1(\beta)+\beta^2 J_1'(\beta)
+\gamma J_1(\gamma)+\gamma^2 J_1'(\gamma)]\}\\ \\

\ds\alpha_{yx}=\frac{4\pi k_B^2Te\tau}{3\hbar^2c}
\frac{t_b^2}{t_a^2}\frac{\obtt[1+2(\obtt)^2]}
{[1+(\obtt)^2]^2} \\ \\

\alpha_{xy}=-\alpha_{yx}\\ \\

\ds\alpha_{zx}=\frac{2\pi k_B^2Te\tau}{3\hbar^2 b}\left\{\gamma^2
\frac{\octt[1+2(\octt)^2]}{[1+(\octt)^2]^2}-\frac{\gamma^3}{4}
\cdot\hall{2\octt}\cdot
\left[J_1(\gamma)\cdot\frac{5+7(2\octt)^2}{1+(2\octt)^2}
+3\gamma J_1'(\gamma)\right]-\frac{\gamma\beta}{2}\cdot
\right. \\ \\ \ds\left.
\left[\beta\jgo+\gamma\jbo\right]\left[\pehh{\obtt-\octt}
+\pehh{\obtt+\octt}\right]\right. \\ \\ \ds\left.
-\frac{3\gamma\beta}{4}\left[\beta\go J_1'\left(\go\right)
+\gamma\bo J_1'\left(\bo\right)\right]\cdot
\left[\hall{\obtt-\octt}+\hall{\obtt+\octt}\right]
\right\}\\ \\

\ds\alpha_{xz}=-\alpha_{zx}\\ \\

\ds\alpha_{yy}=\frac{2\pi k_B^2Te\tau b}{3\hbar^2ac}\frac{t_b^2}{t_a^2}
\frac{1+3(\obtt)^2}{[1+(\obtt)^2]^2}\\ \\

\ds\alpha_{yz}=\frac{\pi k_B^2Te\tau \beta\gamma^2}{3\hbar^2a}\left\{
\jbo\left[\frac{2+3(\obtt-\octt)^2}{[1+(\obtt-\octt)^2]^2}
+\frac{2+3(\obtt+\octt)^2}{[1+(\obtt+\octt)^2]^2}\right]
\right.\\ \\ \qquad \ds\left.
+\frac{3\beta\oct}{2\obt}J_1'\left(\bo\right)
\left[\mr{\obtt-\octt}+\mr{\obtt+\octt}\right]
\right\}\\ \\

\ds\alpha_{zy}=\frac{\pi k_B^2Te\tau\gamma\beta^2}{3\hbar^2a}\left\{
\jgo\left[\frac{2+3(\obtt-\octt)^2}{[1+(\obtt-\octt)^2]^2}
+\frac{2+3(\obtt+\octt)^2}{[1+(\obtt+\octt)^2]^2}\right]
\right.\\ \\ \qquad \ds \left.
+\frac{3\gamma\obt}{2\oct}J_1'\left(\go\right)
\left[\mr{\obtt-\octt}+\mr{\obtt+\octt}\right]
\right\}\\ \\

\ds\alpha_{zz}=\frac{\pi k_B^2Te\tau \gamma^2 c}{\hbar^2ab}\left\{
\frac{1+3(\octt)^2}{[1+(\octt)^2]^2}-\frac{\gamma J_1(\gamma)}{2}\cdot
\frac{7+11\cdot(2\octt)^2}{[1+(2\octt)^2]^2}-\frac{3\gamma^2J_1'(\gamma)}
{1+(2\octt)^2}
\right. \\ \\ \ds\qquad\left.
+\beta\jbo\left[\frac{2+3(\obtt-\octt)^2}{[1+(\obtt-\octt)^2]^2}
-\frac{2+3(\obtt+\octt)^2}{[1+(\obtt+\octt)^2]^2}\right]
\right. \\ \\ \ds\qquad \left.
+3\bo J_1'\left(\bo\right)\left[\mr{\obtt-\octt}-\mr{\obtt+\octt}\right]
\right\}
\end{array}\right.
\ee
\end{widetext}
Once we have $\sigma$ and $\alpha$, we can calculate $\ts{S}=\sigma^{-1}
\cdot\alpha$. The analytic form of $\ts{S}$ would be too long to write
down here since every term in $\ts{S}$ is the sum of three products
of two matrix elements. We just numerically calculate the matrix
product.  Here we only present the angular dependence of all quantities
in order to compare with experiment. In principle one could obtain
temperature and field dependence with these formulae.

\bibliographystyle{apsrev}
\bibliography{organic}

\begin{thebibliography}{42}
\expandafter\ifx\csname natexlab\endcsname\relax\def\natexlab#1{#1}\fi
\expandafter\ifx\csname bibnamefont\endcsname\relax
  \def\bibnamefont#1{#1}\fi
\expandafter\ifx\csname bibfnamefont\endcsname\relax
  \def\bibfnamefont#1{#1}\fi
\expandafter\ifx\csname citenamefont\endcsname\relax
  \def\citenamefont#1{#1}\fi
\expandafter\ifx\csname url\endcsname\relax
  \def\url#1{\texttt{#1}}\fi
\expandafter\ifx\csname urlprefix\endcsname\relax\def\urlprefix{URL }\fi
\providecommand{\bibinfo}[2]{#2}
\providecommand{\eprint}[2][]{\url{#2}}

\bibitem[{\citenamefont{Lee et~al.}(1997)\citenamefont{Lee, Naughton, Danner,
  and Chaikin}}]{Lee1997}
\bibinfo{author}{\bibfnamefont{I.~J.} \bibnamefont{Lee}},
  \bibinfo{author}{\bibfnamefont{M.~J.} \bibnamefont{Naughton}},
  \bibinfo{author}{\bibfnamefont{G.~M.} \bibnamefont{Danner}},
  \bibnamefont{and} \bibinfo{author}{\bibfnamefont{P.~M.}
  \bibnamefont{Chaikin}}, \bibinfo{journal}{Phys. Rev. Lett.}
  \textbf{\bibinfo{volume}{78}}, \bibinfo{pages}{3555} (\bibinfo{year}{1997}).

\bibitem[{\citenamefont{Lee et~al.}(2002{\natexlab{a}})\citenamefont{Lee,
  Brown, Clark, Strouse, Naughton, Kang, and Chaikin}}]{Lee2002K}
\bibinfo{author}{\bibfnamefont{I.~J.} \bibnamefont{Lee}},
  \bibinfo{author}{\bibfnamefont{S.~E.} \bibnamefont{Brown}},
  \bibinfo{author}{\bibfnamefont{W.~G.} \bibnamefont{Clark}},
  \bibinfo{author}{\bibfnamefont{M.~J.} \bibnamefont{Strouse}},
  \bibinfo{author}{\bibfnamefont{M.~J.} \bibnamefont{Naughton}},
  \bibinfo{author}{\bibfnamefont{W.}~\bibnamefont{Kang}}, \bibnamefont{and}
  \bibinfo{author}{\bibfnamefont{P.~M.} \bibnamefont{Chaikin}},
  \bibinfo{journal}{Phys. Rev. Lett.} \textbf{\bibinfo{volume}{88}},
  \bibinfo{pages}{017004} (\bibinfo{year}{2002}{\natexlab{a}}).

\bibitem[{\citenamefont{Lee et~al.}(2002{\natexlab{b}})\citenamefont{Lee,
  Chaikin, and Naughton}}]{Lee2002U}
\bibinfo{author}{\bibfnamefont{I.~J.} \bibnamefont{Lee}},
  \bibinfo{author}{\bibfnamefont{P.~M.} \bibnamefont{Chaikin}},
  \bibnamefont{and} \bibinfo{author}{\bibfnamefont{M.~J.}
  \bibnamefont{Naughton}}, \bibinfo{journal}{Phys. Rev. Lett.}
  \textbf{\bibinfo{volume}{88}}, \bibinfo{pages}{207002}
  (\bibinfo{year}{2002}{\natexlab{b}}).

\bibitem[{\citenamefont{Ishiguro et~al.}(1997)\citenamefont{Ishiguro, Yamaji,
  and Saito}}]{Ishiguro1997}
\bibinfo{author}{\bibfnamefont{T.}~\bibnamefont{Ishiguro}},
  \bibinfo{author}{\bibfnamefont{K.}~\bibnamefont{Yamaji}}, \bibnamefont{and}
  \bibinfo{author}{\bibfnamefont{G.}~\bibnamefont{Saito}},
  \emph{\bibinfo{title}{Organic Superconductor}}
  (\bibinfo{publisher}{Springer}, \bibinfo{year}{1997}).

\bibitem[{\citenamefont{Osada et~al.}(1991)\citenamefont{Osada, Kawasumi,
  Kagoshima, Miura, and Saito}}]{Osada1991}
\bibinfo{author}{\bibfnamefont{T.}~\bibnamefont{Osada}},
  \bibinfo{author}{\bibfnamefont{A.}~\bibnamefont{Kawasumi}},
  \bibinfo{author}{\bibfnamefont{S.}~\bibnamefont{Kagoshima}},
  \bibinfo{author}{\bibfnamefont{N.}~\bibnamefont{Miura}}, \bibnamefont{and}
  \bibinfo{author}{\bibfnamefont{G.}~\bibnamefont{Saito}},
  \bibinfo{journal}{Phys. Rev. Lett.} \textbf{\bibinfo{volume}{66}},
  \bibinfo{pages}{1525} (\bibinfo{year}{1991}).

\bibitem[{\citenamefont{Naughton et~al.}(1991)\citenamefont{Naughton, Chung,
  Chaparala, Bu, and Coppens}}]{Naughton1991}
\bibinfo{author}{\bibfnamefont{M.~J.} \bibnamefont{Naughton}},
  \bibinfo{author}{\bibfnamefont{O.~H.} \bibnamefont{Chung}},
  \bibinfo{author}{\bibfnamefont{M.}~\bibnamefont{Chaparala}},
  \bibinfo{author}{\bibfnamefont{X.}~\bibnamefont{Bu}}, \bibnamefont{and}
  \bibinfo{author}{\bibfnamefont{P.}~\bibnamefont{Coppens}},
  \bibinfo{journal}{Phys. Rev. Lett.} \textbf{\bibinfo{volume}{67}},
  \bibinfo{pages}{3712} (\bibinfo{year}{1991}).

\bibitem[{\citenamefont{Kang et~al.}(1992)\citenamefont{Kang, Hannahs, and
  Chaikin}}]{Kang1992MA}
\bibinfo{author}{\bibfnamefont{W.}~\bibnamefont{Kang}},
  \bibinfo{author}{\bibfnamefont{S.~T.} \bibnamefont{Hannahs}},
  \bibnamefont{and} \bibinfo{author}{\bibfnamefont{P.~M.}
  \bibnamefont{Chaikin}}, \bibinfo{journal}{Phys. Rev. Lett.}
  \textbf{\bibinfo{volume}{69}}, \bibinfo{pages}{2827} (\bibinfo{year}{1992}).

\bibitem[{\citenamefont{Lebed}(1986)}]{Lebed1986}
\bibinfo{author}{\bibfnamefont{A.~G.} \bibnamefont{Lebed}},
  \bibinfo{journal}{JETP} \textbf{\bibinfo{volume}{43}}, \bibinfo{pages}{137}
  (\bibinfo{year}{1986}).

\bibitem[{\citenamefont{Lebed and Bak}(1989)}]{Lebed1989}
\bibinfo{author}{\bibfnamefont{A.~G.} \bibnamefont{Lebed}} \bibnamefont{and}
  \bibinfo{author}{\bibfnamefont{P.}~\bibnamefont{Bak}},
  \bibinfo{journal}{Phys. Rev. Lett.} \textbf{\bibinfo{volume}{63}},
  \bibinfo{pages}{1315} (\bibinfo{year}{1989}).

\bibitem[{\citenamefont{Strong et~al.}(1994)\citenamefont{Strong, Clarke, and
  Anderson}}]{Strong1994}
\bibinfo{author}{\bibfnamefont{S.~P.} \bibnamefont{Strong}},
  \bibinfo{author}{\bibfnamefont{D.~G.} \bibnamefont{Clarke}},
  \bibnamefont{and} \bibinfo{author}{\bibfnamefont{P.~W.}
  \bibnamefont{Anderson}}, \bibinfo{journal}{Phys. Rev. Lett.}
  \textbf{\bibinfo{volume}{73}}, \bibinfo{pages}{1007} (\bibinfo{year}{1994}).

\bibitem[{\citenamefont{Osada et~al.}(1992)\citenamefont{Osada, Kagoshima, and
  Miura}}]{Osada1992}
\bibinfo{author}{\bibfnamefont{T.}~\bibnamefont{Osada}},
  \bibinfo{author}{\bibfnamefont{S.}~\bibnamefont{Kagoshima}},
  \bibnamefont{and} \bibinfo{author}{\bibfnamefont{N.}~\bibnamefont{Miura}},
  \bibinfo{journal}{Phys. Rev. B} \textbf{\bibinfo{volume}{46}},
  \bibinfo{pages}{1812} (\bibinfo{year}{1992}).

\bibitem[{\citenamefont{Osada}(1998)}]{Osada1998}
\bibinfo{author}{\bibfnamefont{T.}~\bibnamefont{Osada}},
  \bibinfo{journal}{Physica B} \textbf{\bibinfo{volume}{256}},
  \bibinfo{pages}{633} (\bibinfo{year}{1998}).

\bibitem[{\citenamefont{Maki}(1992)}]{Maki1992}
\bibinfo{author}{\bibfnamefont{K.}~\bibnamefont{Maki}}, \bibinfo{journal}{Phys.
  Rev. B} \textbf{\bibinfo{volume}{45}}, \bibinfo{pages}{R5111}
  (\bibinfo{year}{1992}).

\bibitem[{\citenamefont{Yakovenko}(1992)}]{Yakovenko1992}
\bibinfo{author}{\bibfnamefont{V.~M.} \bibnamefont{Yakovenko}},
  \bibinfo{journal}{Phys. Rev. Lett.} \textbf{\bibinfo{volume}{68}},
  \bibinfo{pages}{3607} (\bibinfo{year}{1992}).

\bibitem[{\citenamefont{Chaikin}(1992)}]{Chaikin1992}
\bibinfo{author}{\bibfnamefont{P.~M.} \bibnamefont{Chaikin}},
  \bibinfo{journal}{Phys. Rev. Lett.} \textbf{\bibinfo{volume}{69}},
  \bibinfo{pages}{2831} (\bibinfo{year}{1992}).

\bibitem[{\citenamefont{Lebed}(1996)}]{Lebed1996}
\bibinfo{author}{\bibfnamefont{A.~G.} \bibnamefont{Lebed}},
  \bibinfo{journal}{J. Phys. I (France)} \textbf{\bibinfo{volume}{6}},
  \bibinfo{pages}{1819} (\bibinfo{year}{1996}).

\bibitem[{\citenamefont{Ong et~al.}(2004)\citenamefont{Ong, Wu, Chaikin, and
  Anderson}}]{Ong2004}
\bibinfo{author}{\bibfnamefont{N.~P.} \bibnamefont{Ong}},
  \bibinfo{author}{\bibfnamefont{W.}~\bibnamefont{Wu}},
  \bibinfo{author}{\bibfnamefont{P.~M.} \bibnamefont{Chaikin}},
  \bibnamefont{and} \bibinfo{author}{\bibfnamefont{P.~W.}
  \bibnamefont{Anderson}}, \bibinfo{journal}{EuroPhys. Lett.}
  \textbf{\bibinfo{volume}{66}}, \bibinfo{pages}{579} (\bibinfo{year}{2004}).

\bibitem[{\citenamefont{Wu et~al.}(2003)\citenamefont{Wu, Lee, and
  Chaikin}}]{Wu2003}
\bibinfo{author}{\bibfnamefont{W.}~\bibnamefont{Wu}},
  \bibinfo{author}{\bibfnamefont{I.~J.} \bibnamefont{Lee}}, \bibnamefont{and}
  \bibinfo{author}{\bibfnamefont{P.~M.} \bibnamefont{Chaikin}},
  \bibinfo{journal}{Phys. Rev. Lett.} \textbf{\bibinfo{volume}{91}},
  \bibinfo{pages}{056601} (\bibinfo{year}{2003}).

\bibitem[{\citenamefont{Choi et~al.}(2005)\citenamefont{Choi, Brooks, Kang, Jo,
  and Kang}}]{Choi2005}
\bibinfo{author}{\bibfnamefont{E.~S.} \bibnamefont{Choi}},
  \bibinfo{author}{\bibfnamefont{J.~S.} \bibnamefont{Brooks}},
  \bibinfo{author}{\bibfnamefont{H.}~\bibnamefont{Kang}},
  \bibinfo{author}{\bibfnamefont{Y.~J.} \bibnamefont{Jo}}, \bibnamefont{and}
  \bibinfo{author}{\bibfnamefont{W.}~\bibnamefont{Kang}},
  \textbf{\bibinfo{volume}{cond-mat/0501649}} (\bibinfo{year}{2005}).

\bibitem[{\citenamefont{Wang et~al.}(2001)\citenamefont{Wang, Xu, Kakeshita,
  Uchida, Ono, Ando, and Ong}}]{Wang2001}
\bibinfo{author}{\bibfnamefont{Y.}~\bibnamefont{Wang}},
  \bibinfo{author}{\bibfnamefont{Z.~A.} \bibnamefont{Xu}},
  \bibinfo{author}{\bibfnamefont{T.}~\bibnamefont{Kakeshita}},
  \bibinfo{author}{\bibfnamefont{S.}~\bibnamefont{Uchida}},
  \bibinfo{author}{\bibfnamefont{S.}~\bibnamefont{Ono}},
  \bibinfo{author}{\bibfnamefont{Y.}~\bibnamefont{Ando}}, \bibnamefont{and}
  \bibinfo{author}{\bibfnamefont{N.~P.} \bibnamefont{Ong}},
  \bibinfo{journal}{Phys. Rev. B} \textbf{\bibinfo{volume}{64}},
  \bibinfo{pages}{224519} (\bibinfo{year}{2001}).

\bibitem[{\citenamefont{Wang et~al.}(2002{\natexlab{a}})\citenamefont{Wang,
  Ong, Xu, Kakeshita, Uchida, Bonn, Liang, and Hardy}}]{Wang2002p}
\bibinfo{author}{\bibfnamefont{Y.}~\bibnamefont{Wang}},
  \bibinfo{author}{\bibfnamefont{N.~P.} \bibnamefont{Ong}},
  \bibinfo{author}{\bibfnamefont{Z.~A.} \bibnamefont{Xu}},
  \bibinfo{author}{\bibfnamefont{T.}~\bibnamefont{Kakeshita}},
  \bibinfo{author}{\bibfnamefont{S.}~\bibnamefont{Uchida}},
  \bibinfo{author}{\bibfnamefont{D.~A.} \bibnamefont{Bonn}},
  \bibinfo{author}{\bibfnamefont{R.}~\bibnamefont{Liang}}, \bibnamefont{and}
  \bibinfo{author}{\bibfnamefont{W.~N.} \bibnamefont{Hardy}},
  \bibinfo{journal}{Phys. Rev. Lett.} \textbf{\bibinfo{volume}{88}},
  \bibinfo{pages}{257003} (\bibinfo{year}{2002}{\natexlab{a}}).

\bibitem[{\citenamefont{Wang et~al.}(2002{\natexlab{b}})\citenamefont{Wang,
  Ono, Onose, Gu, Ando, Tokura, Uchida, and Ong}}]{Wang2002s}
\bibinfo{author}{\bibfnamefont{Y.}~\bibnamefont{Wang}},
  \bibinfo{author}{\bibfnamefont{S.}~\bibnamefont{Ono}},
  \bibinfo{author}{\bibfnamefont{Y.}~\bibnamefont{Onose}},
  \bibinfo{author}{\bibfnamefont{G.}~\bibnamefont{Gu}},
  \bibinfo{author}{\bibfnamefont{Y.}~\bibnamefont{Ando}},
  \bibinfo{author}{\bibfnamefont{Y.}~\bibnamefont{Tokura}},
  \bibinfo{author}{\bibfnamefont{S.}~\bibnamefont{Uchida}}, \bibnamefont{and}
  \bibinfo{author}{\bibfnamefont{N.~P.} \bibnamefont{Ong}},
  \bibinfo{journal}{Science} \textbf{\bibinfo{volume}{299}},
  \bibinfo{pages}{86} (\bibinfo{year}{2002}{\natexlab{b}}).

\bibitem[{\citenamefont{Wu et~al.}(2005)\citenamefont{Wu, Chaikin, Kang,
  Shinagawa, Yu, and Brown}}]{Wu2005}
\bibinfo{author}{\bibfnamefont{W.}~\bibnamefont{Wu}},
  \bibinfo{author}{\bibfnamefont{P.~M.} \bibnamefont{Chaikin}},
  \bibinfo{author}{\bibfnamefont{W.}~\bibnamefont{Kang}},
  \bibinfo{author}{\bibfnamefont{J.}~\bibnamefont{Shinagawa}},
  \bibinfo{author}{\bibfnamefont{W.}~\bibnamefont{Yu}}, \bibnamefont{and}
  \bibinfo{author}{\bibfnamefont{S.~E.} \bibnamefont{Brown}},
  \bibinfo{journal}{Phys. Rev. Lett.} \textbf{\bibinfo{volume}{94}},
  \bibinfo{pages}{097004} (\bibinfo{year}{2005}).

\bibitem[{\citenamefont{Vescoli et~al.}(1998)\citenamefont{Vescoli, Degiorgi,
  Henderson, Grüner, Starkey, and Montgomery}}]{Vescoli1998}
\bibinfo{author}{\bibfnamefont{V.}~\bibnamefont{Vescoli}},
  \bibinfo{author}{\bibfnamefont{L.}~\bibnamefont{Degiorgi}},
  \bibinfo{author}{\bibfnamefont{W.}~\bibnamefont{Henderson}},
  \bibinfo{author}{\bibfnamefont{G.}~\bibnamefont{Grüner}},
  \bibinfo{author}{\bibfnamefont{K.~P.} \bibnamefont{Starkey}},
  \bibnamefont{and} \bibinfo{author}{\bibfnamefont{L.~K.}
  \bibnamefont{Montgomery}}, \bibinfo{journal}{Science}
  \textbf{\bibinfo{volume}{281}}, \bibinfo{pages}{1181} (\bibinfo{year}{1998}).

\bibitem[{\citenamefont{Lorenz et~al.}(2002)\citenamefont{Lorenz, Hofmann,
  M.Gruninger, Freimuth, Uhrig, Dumm, and Dressel}}]{Lorenz2002}
\bibinfo{author}{\bibfnamefont{T.}~\bibnamefont{Lorenz}},
  \bibinfo{author}{\bibfnamefont{M.}~\bibnamefont{Hofmann}},
  \bibinfo{author}{\bibnamefont{M.Gruninger}},
  \bibinfo{author}{\bibfnamefont{A.}~\bibnamefont{Freimuth}},
  \bibinfo{author}{\bibfnamefont{G.~S.} \bibnamefont{Uhrig}},
  \bibinfo{author}{\bibfnamefont{M.}~\bibnamefont{Dumm}}, \bibnamefont{and}
  \bibinfo{author}{\bibfnamefont{M.}~\bibnamefont{Dressel}},
  \bibinfo{journal}{Nature} \textbf{\bibinfo{volume}{418}},
  \bibinfo{pages}{614} (\bibinfo{year}{2002}).

\bibitem[{\citenamefont{Yu}(1990)}]{Yu_thesis1990}
\bibinfo{author}{\bibfnamefont{R.}~\bibnamefont{Yu}}, Ph.D. thesis,
  \bibinfo{school}{U. Penn.} (\bibinfo{year}{1990}).

\bibitem[{\citenamefont{Wu}(2004)}]{Wu_thesis2004}
\bibinfo{author}{\bibfnamefont{W.}~\bibnamefont{Wu}}, Ph.D. thesis,
  \bibinfo{school}{Princeton} (\bibinfo{year}{2004}).

\bibitem[{\citenamefont{Josephson}(1962)}]{Josephson1962}
\bibinfo{author}{\bibfnamefont{B.~D.} \bibnamefont{Josephson}},
  \bibinfo{journal}{Phys. Lett.} \textbf{\bibinfo{volume}{1}},
  \bibinfo{pages}{251} (\bibinfo{year}{1962}).

\bibitem[{\citenamefont{Josephson}(1965)}]{Josephson1965}
\bibinfo{author}{\bibfnamefont{B.~D.} \bibnamefont{Josephson}},
  \bibinfo{journal}{Adv. Phys.} \textbf{\bibinfo{volume}{14}},
  \bibinfo{pages}{419} (\bibinfo{year}{1965}).

\bibitem[{\citenamefont{Kang et~al.}(1993)\citenamefont{Kang, Hannahs, and
  Chaikin}}]{Kang1993}
\bibinfo{author}{\bibfnamefont{W.}~\bibnamefont{Kang}},
  \bibinfo{author}{\bibfnamefont{S.~T.} \bibnamefont{Hannahs}},
  \bibnamefont{and} \bibinfo{author}{\bibfnamefont{P.~M.}
  \bibnamefont{Chaikin}}, \bibinfo{journal}{Phys. Rev. Lett.}
  \textbf{\bibinfo{volume}{70}}, \bibinfo{pages}{3091} (\bibinfo{year}{1993}).

\bibitem[{\citenamefont{Danner et~al.}(1994)\citenamefont{Danner, Kang, and
  Chaikin}}]{Danner1994}
\bibinfo{author}{\bibfnamefont{G.~M.} \bibnamefont{Danner}},
  \bibinfo{author}{\bibfnamefont{W.}~\bibnamefont{Kang}}, \bibnamefont{and}
  \bibinfo{author}{\bibfnamefont{P.~M.} \bibnamefont{Chaikin}},
  \bibinfo{journal}{Phys. Rev. Lett.} \textbf{\bibinfo{volume}{72}},
  \bibinfo{pages}{3714} (\bibinfo{year}{1994}).

\bibitem[{\citenamefont{Osada et~al.}(1996)\citenamefont{Osada, Kagoshima, and
  Miura}}]{Osada1996}
\bibinfo{author}{\bibfnamefont{T.}~\bibnamefont{Osada}},
  \bibinfo{author}{\bibfnamefont{S.}~\bibnamefont{Kagoshima}},
  \bibnamefont{and} \bibinfo{author}{\bibfnamefont{N.}~\bibnamefont{Miura}},
  \bibinfo{journal}{Phys. Rev. Lett.} \textbf{\bibinfo{volume}{77}},
  \bibinfo{pages}{5261} (\bibinfo{year}{1996}).

\bibitem[{\citenamefont{Lee and Naughton}(1998)}]{Lee1998a}
\bibinfo{author}{\bibfnamefont{I.~J.} \bibnamefont{Lee}} \bibnamefont{and}
  \bibinfo{author}{\bibfnamefont{M.~J.} \bibnamefont{Naughton}},
  \bibinfo{journal}{Phys. Rev. B} \textbf{\bibinfo{volume}{57}},
  \bibinfo{pages}{7423} (\bibinfo{year}{1998}).

\bibitem[{\citenamefont{Onsager}(1931{\natexlab{a}})}]{Onsager1931a}
\bibinfo{author}{\bibfnamefont{L.}~\bibnamefont{Onsager}},
  \bibinfo{journal}{Phys. Rev.} \textbf{\bibinfo{volume}{37}},
  \bibinfo{pages}{405} (\bibinfo{year}{1931}{\natexlab{a}}).

\bibitem[{\citenamefont{Onsager}(1931{\natexlab{b}})}]{Onsager1931b}
\bibinfo{author}{\bibfnamefont{L.}~\bibnamefont{Onsager}},
  \bibinfo{journal}{Phys. Rev.} \textbf{\bibinfo{volume}{38}},
  \bibinfo{pages}{2265} (\bibinfo{year}{1931}{\natexlab{b}}).

\bibitem[{\citenamefont{Ashcroft and Mermin}(1976)}]{mermin}
\bibinfo{author}{\bibfnamefont{N.~W.} \bibnamefont{Ashcroft}} \bibnamefont{and}
  \bibinfo{author}{\bibfnamefont{N.~D.} \bibnamefont{Mermin}},
  \emph{\bibinfo{title}{Solid State Physics}} (\bibinfo{publisher}{Saunders},
  \bibinfo{address}{Philadelphia}, \bibinfo{year}{1976}).

\bibitem[{\citenamefont{Press et~al.}(1991)\citenamefont{Press, Flannery,
  Teukolsky, and Vetterling}}]{Recipes}
\bibinfo{author}{\bibfnamefont{W.~H.} \bibnamefont{Press}},
  \bibinfo{author}{\bibfnamefont{B.~P.} \bibnamefont{Flannery}},
  \bibinfo{author}{\bibfnamefont{S.~A.} \bibnamefont{Teukolsky}},
  \bibnamefont{and} \bibinfo{author}{\bibfnamefont{W.~T.}
  \bibnamefont{Vetterling}}, \emph{\bibinfo{title}{Numerical Recipes in Pascal:
  The Art of Scientific Computing}} (\bibinfo{publisher}{Cambridge University
  Press}, \bibinfo{year}{1991}), \bibinfo{edition}{1st} ed.

\bibitem[{\citenamefont{Lebed et~al.}(2004)\citenamefont{Lebed, Bagmet, and
  Naughton}}]{Lebed2004}
\bibinfo{author}{\bibfnamefont{A.~G.} \bibnamefont{Lebed}},
  \bibinfo{author}{\bibfnamefont{N.~N.} \bibnamefont{Bagmet}},
  \bibnamefont{and} \bibinfo{author}{\bibfnamefont{M.~J.}
  \bibnamefont{Naughton}}, \bibinfo{journal}{Phys. Rev. Lett.}
  \textbf{\bibinfo{volume}{93}}, \bibinfo{pages}{157006}
  (\bibinfo{year}{2004}).

\bibitem[{\citenamefont{Osada et~al.}(1999)\citenamefont{Osada, Kami, Kondo,
  and Kagoshima}}]{Osada1999}
\bibinfo{author}{\bibfnamefont{T.}~\bibnamefont{Osada}},
  \bibinfo{author}{\bibfnamefont{N.}~\bibnamefont{Kami}},
  \bibinfo{author}{\bibfnamefont{R.}~\bibnamefont{Kondo}}, \bibnamefont{and}
  \bibinfo{author}{\bibfnamefont{S.}~\bibnamefont{Kagoshima}},
  \bibinfo{journal}{Syn. Met.} \textbf{\bibinfo{volume}{103}},
  \bibinfo{pages}{2024} (\bibinfo{year}{1999}).

\bibitem[{\citenamefont{Nam et~al.}(2005)\citenamefont{Nam, Ardavan, Wu, and
  Chaikin}}]{Nam2005}
\bibinfo{author}{\bibfnamefont{M.-S.} \bibnamefont{Nam}},
  \bibinfo{author}{\bibfnamefont{A.}~\bibnamefont{Ardavan}},
  \bibinfo{author}{\bibfnamefont{W.}~\bibnamefont{Wu}}, \bibnamefont{and}
  \bibinfo{author}{\bibfnamefont{P.~M.} \bibnamefont{Chaikin}}
  (\bibinfo{year}{2005}), \bibinfo{note}{unpublished}.

\bibitem[{\citenamefont{Danner and Chaikin}(1995)}]{Danner1995}
\bibinfo{author}{\bibfnamefont{G.~M.} \bibnamefont{Danner}} \bibnamefont{and}
  \bibinfo{author}{\bibfnamefont{P.~M.} \bibnamefont{Chaikin}},
  \bibinfo{journal}{Phys. Rev. Lett.} \textbf{\bibinfo{volume}{75}},
  \bibinfo{pages}{4690} (\bibinfo{year}{1995}).

\bibitem[{\citenamefont{Chasheshikina}(1998)}]{Katya_thesis1998}
\bibinfo{author}{\bibfnamefont{E.~I.} \bibnamefont{Chasheshikina}}, Ph.D.
  thesis, \bibinfo{school}{Princeton} (\bibinfo{year}{1998}).

\end{thebibliography}

\end{document}